\documentclass{arxiv}
\usepackage{color,ifpdf,latexsym,graphicx,subfig,url,wrapfig}
\title{Fun with Fonts: Algorithmic Typography\tnoteref{t1}}
\tnotetext[t1]{A preliminary version of this paper appeared in
  \emph{Proceedings of the 7th International Conference on Fun with Algorithms}
  (FUN 2014).}
\author[MIT]{Erik D. Demaine}
\ead{edemaine@mit.edu}
\author[MIT]{Martin L. Demaine}
\ead{mdemaine@mit.edu}
\address[MIT]{MIT CSAIL, 32 Vassar St., Cambridge, MA 02139}

\newif\ifabstract
\abstracttrue
\newif\iffull
\ifabstract \fullfalse \else \fulltrue \fi

\let\realmaketitle=\maketitle
\def\maketitle{%
  \bgroup\def\stepcounter##1{\addtocounter{##1}1}\realmaketitle\egroup}

\usepackage
  [breaklinks,bookmarks,bookmarksnumbered,bookmarksopen,bookmarksopenlevel=2]
  {hyperref}
{\makeatletter \hypersetup{pdftitle={\@title}}}
{\makeatletter
 \gdef\xxxmark{%
   \expandafter\ifx\csname @mpargs\endcsname\relax % in minipage?
     \expandafter\ifx\csname @captype\endcsname\relax % in figure/caption?
       \marginpar{xxx}% not in a caption or minipage, can use marginpar
     \else
       xxx % notice trailing space
     \fi
   \else
     xxx % notice trailing space
   \fi}
 \gdef\xxx{\@ifnextchar[\xxx@lab\xxx@nolab}
 \long\gdef\xxx@lab[#1]#2{\textbf{[\xxxmark #2 ---{\sc #1}]}}
 \long\gdef\xxx@nolab#1{\textbf{[\xxxmark #1]}}
}

{\makeatletter \gdef\fps@figure{!tbp}}

\let\realbfseries=\bfseries
\def\bfseries{\realbfseries\boldmath}

\makeatletter
\let\c@lemma=\c@theorem
\let\c@corollary=\c@theorem
\let\c@fact=\c@theorem
\makeatother

\let\realendproof=\endproof
\def\endproof{\hspace*{\fill}$\Box$\realendproof}

\let\epsilon=\varepsilon

\begin{document}

\begin{abstract}
% We present five typefaces we have designed over the years
  Over the past decade, we have designed six typefaces
  based on mathematical theorems and open problems,
  specifically computational geometry.
  These typefaces expose the general public in a unique way to intriguing
  results and hard problems in hinged dissections, geometric tours,
  origami design, computer-aided glass design, physical simulation,
  and protein folding.
  In particular, most of these typefaces include puzzle fonts,
  where reading the intended message requires solving a series of puzzles
  which illustrate the challenge of the underlying algorithmic problem.
\end{abstract}

\begin{keyword}
  puzzles, typefaces, art, design
\end{keyword}

\maketitle

\section{Introduction}

Scientists use fonts every day to express their research
through the written word.  But what if the font itself communicated
(the spirit of) the research?  What if the way text is written,
and not just the text itself, engages the reader in the science?

We have been designing a series of typefaces (font families) based on our
computational geometry research.
%, both completed theorems and partially understood open problems.
They are \emph{mathematical typefaces} and \emph{algorithmic typefaces}
in the sense that they illustrate mathematical and algorithmic structures,
theorems, and/or open problems.
In all but one family, we include \emph{puzzle typefaces}
where reading the text itself requires engaging with those
same mathematical structures.  With a careful combination of puzzle
and nonpuzzle variants, these typefaces enable the general public
to explore the underlying mathematical structures and appreciate their
inherent beauty, challenge, and fun.

\begin{figure}[t]
  \centering

  \includegraphics[width=0.33\linewidth]{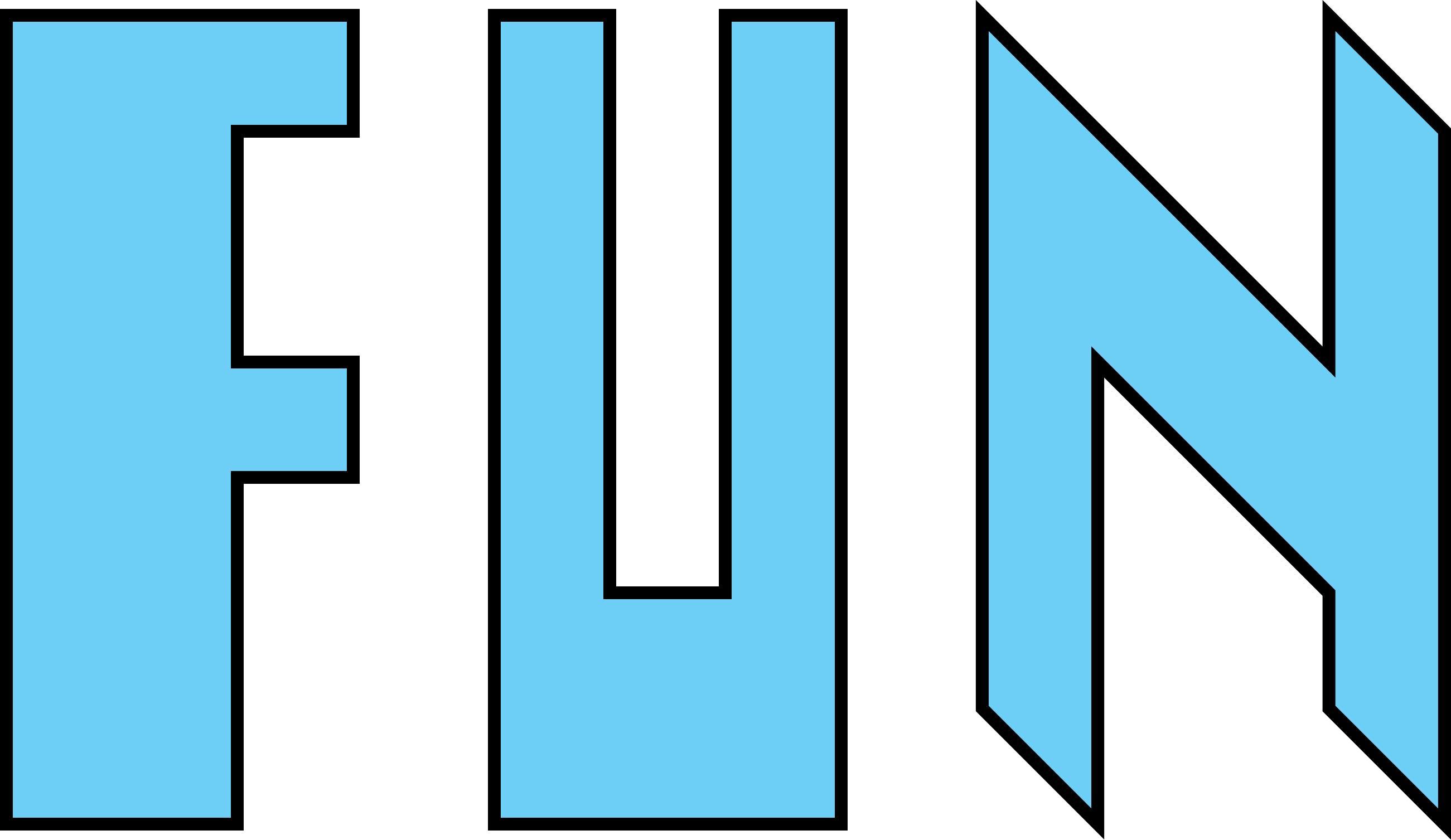}

  \medskip
  \includegraphics[width=0.45\linewidth]{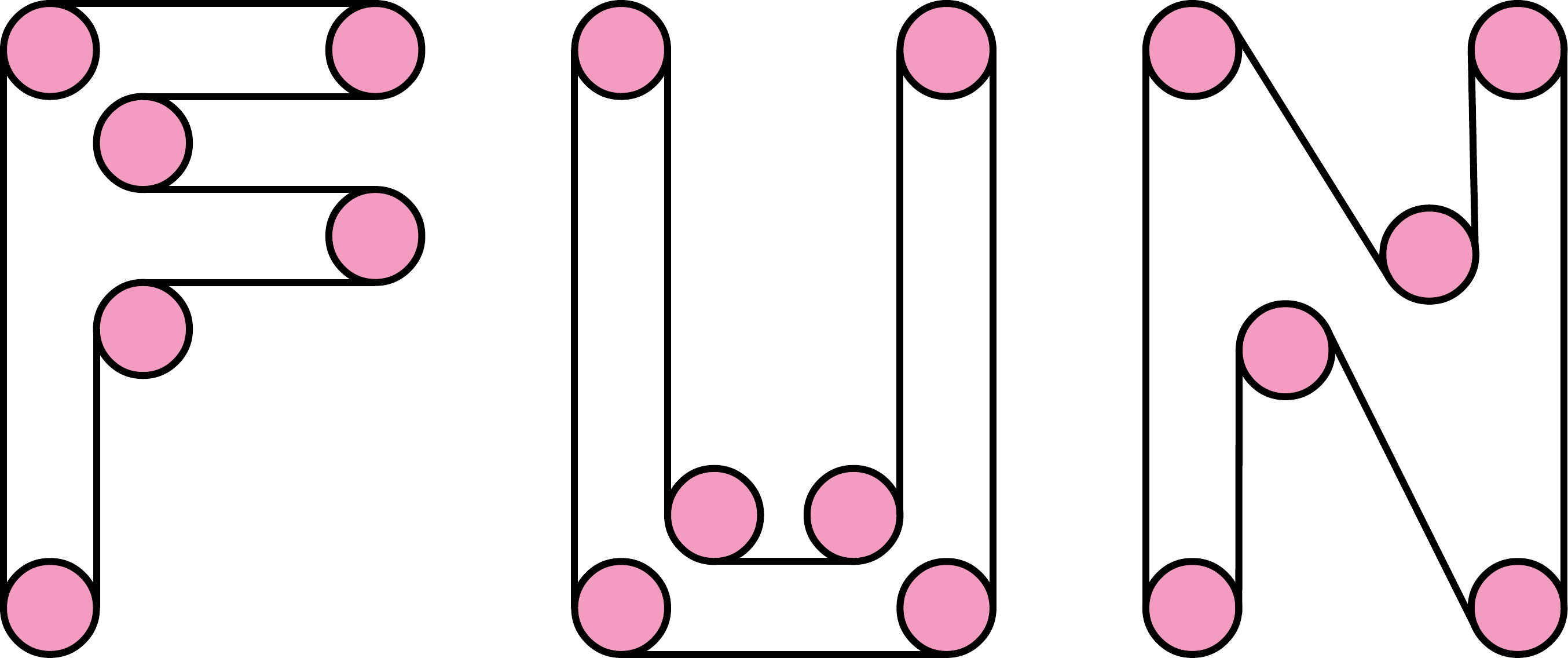}\hfill
  \includegraphics[width=0.45\linewidth]{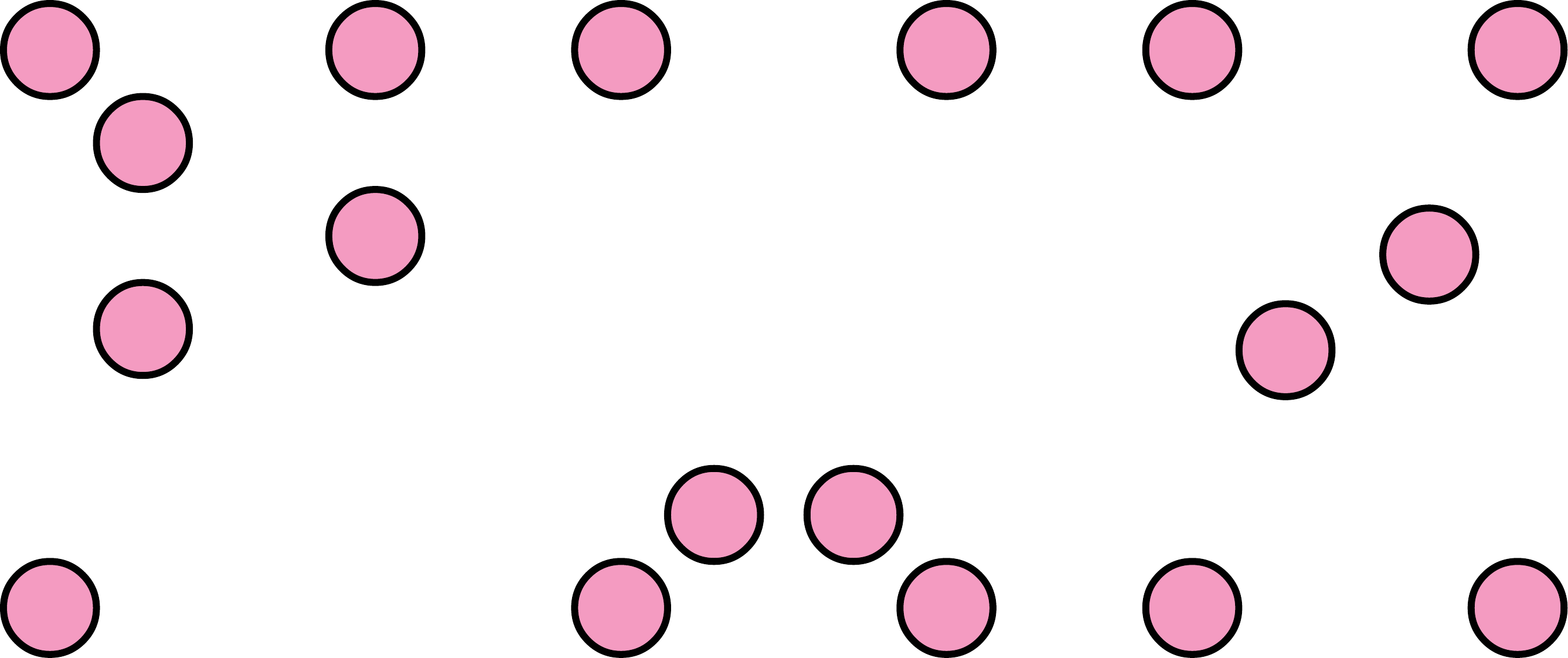}

  \medskip
  \includegraphics[width=0.3\linewidth]{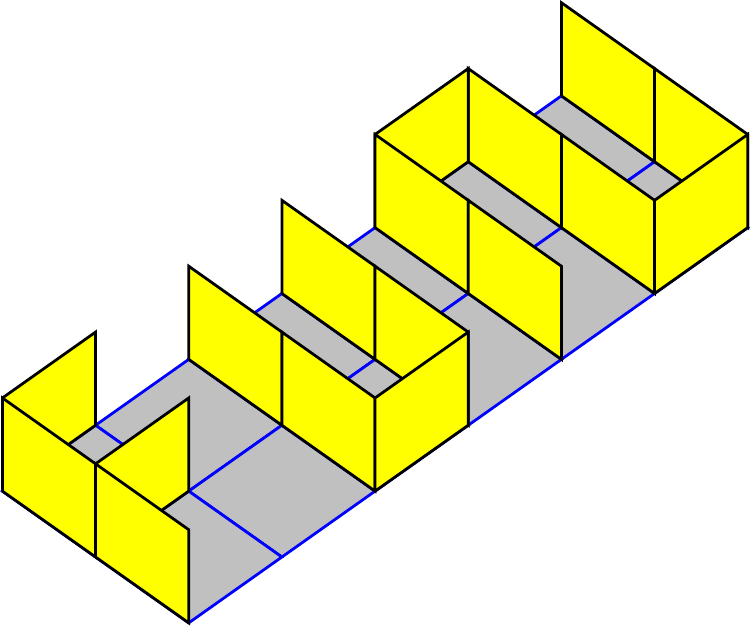}\hfill
  \includegraphics[width=0.6\linewidth]{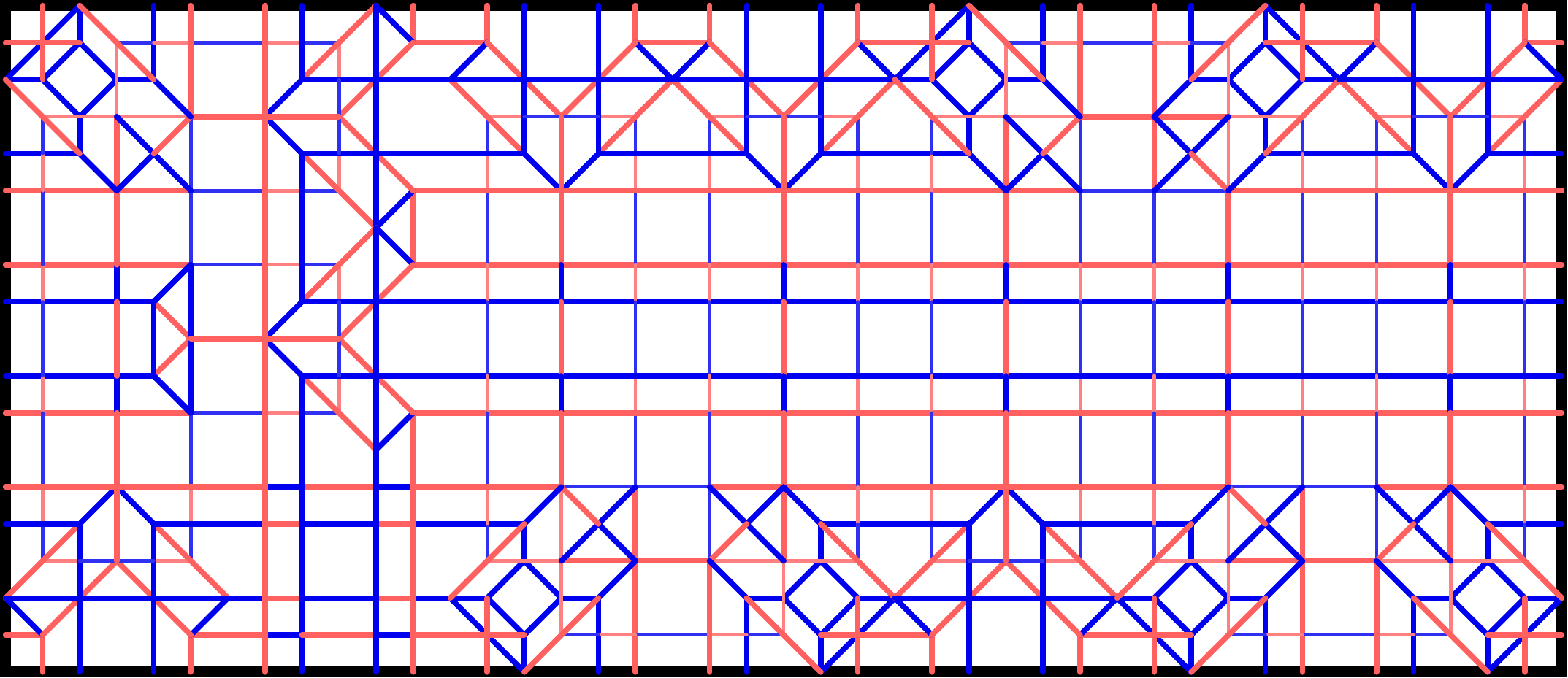}

  \medskip
  \begin{tabular}{cc}
    \includegraphics[height=0.3in]{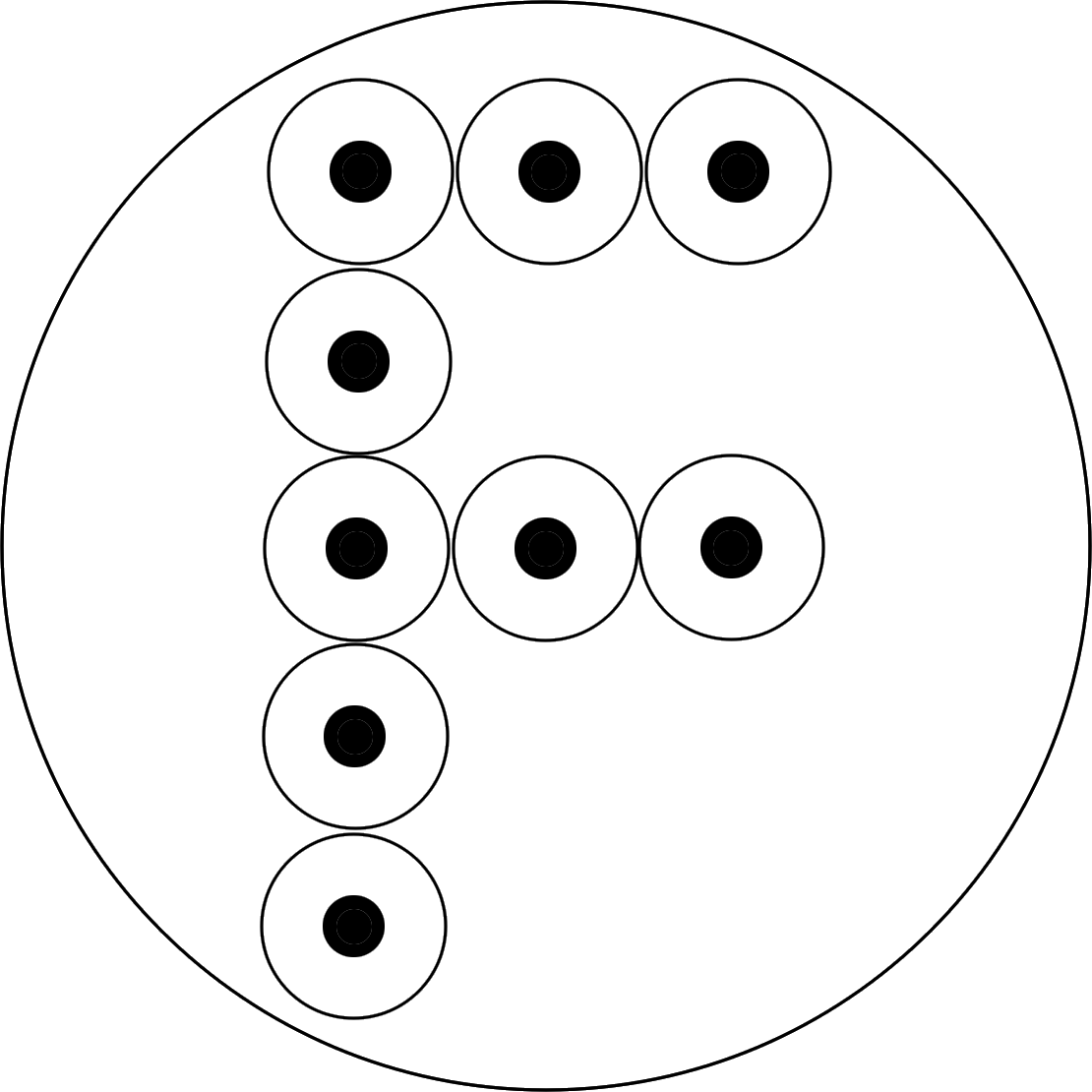} &
    \includegraphics[angle=90,height=0.3in]{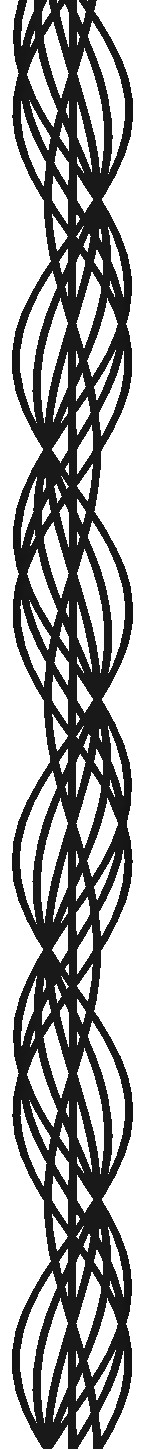} \\
    \includegraphics[height=0.3in]{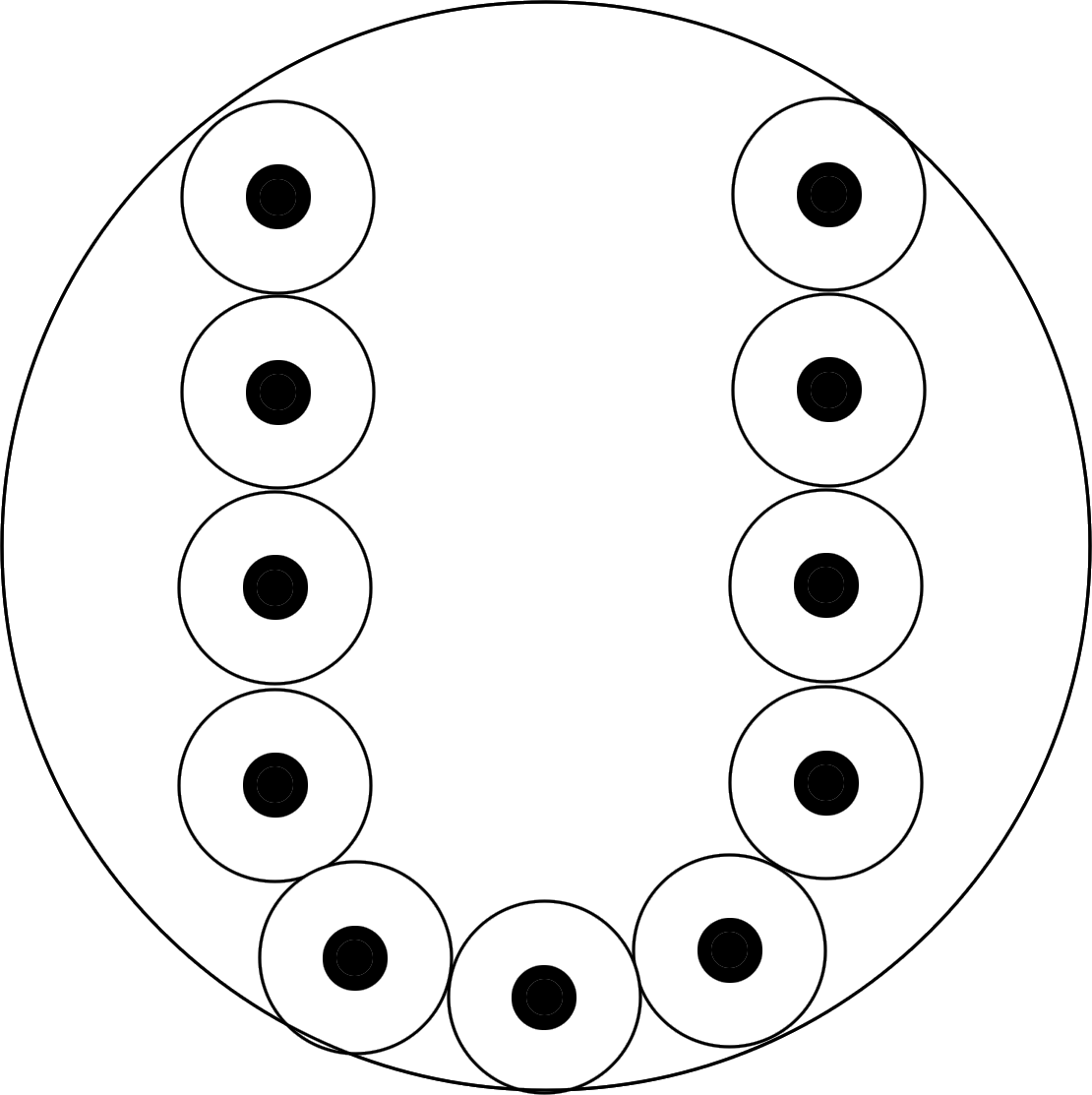} &
    \includegraphics[angle=90,height=0.3in]{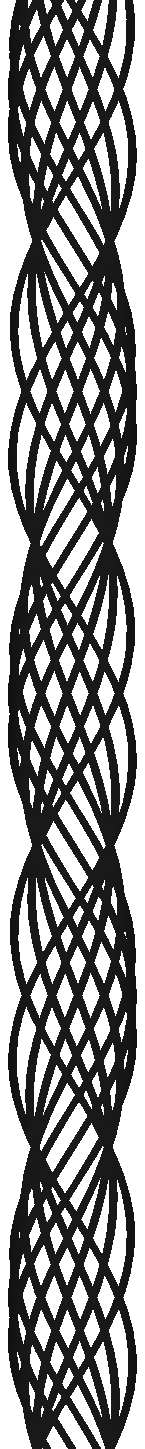} \\
    \includegraphics[height=0.3in]{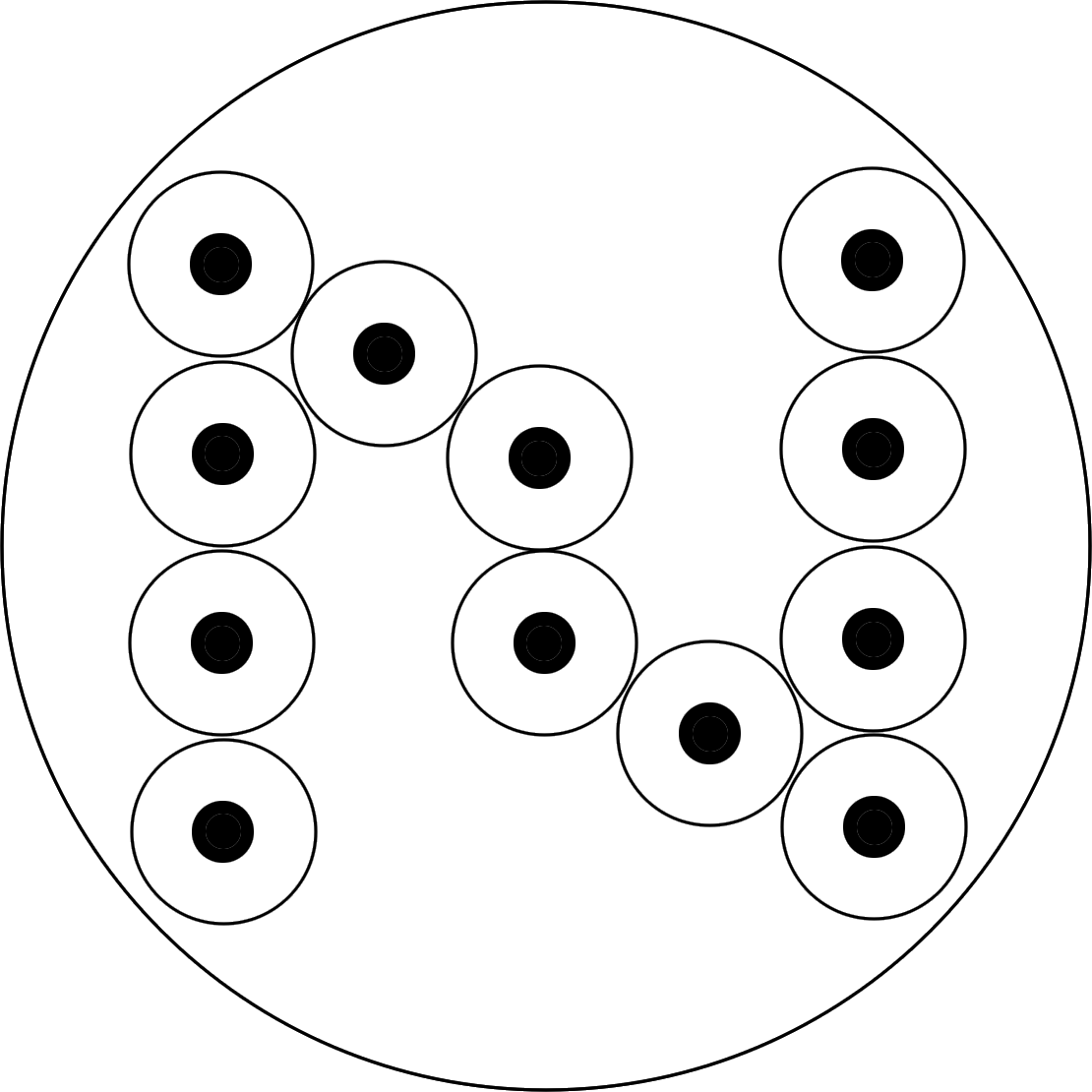} &
    \includegraphics[angle=90,height=0.3in]{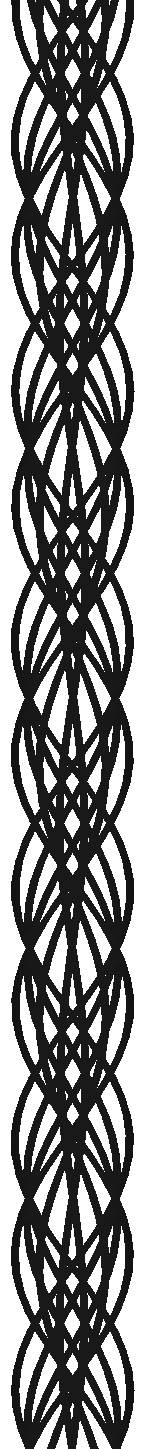}
  \end{tabular}

  \medskip
  \includegraphics[width=0.4\linewidth]{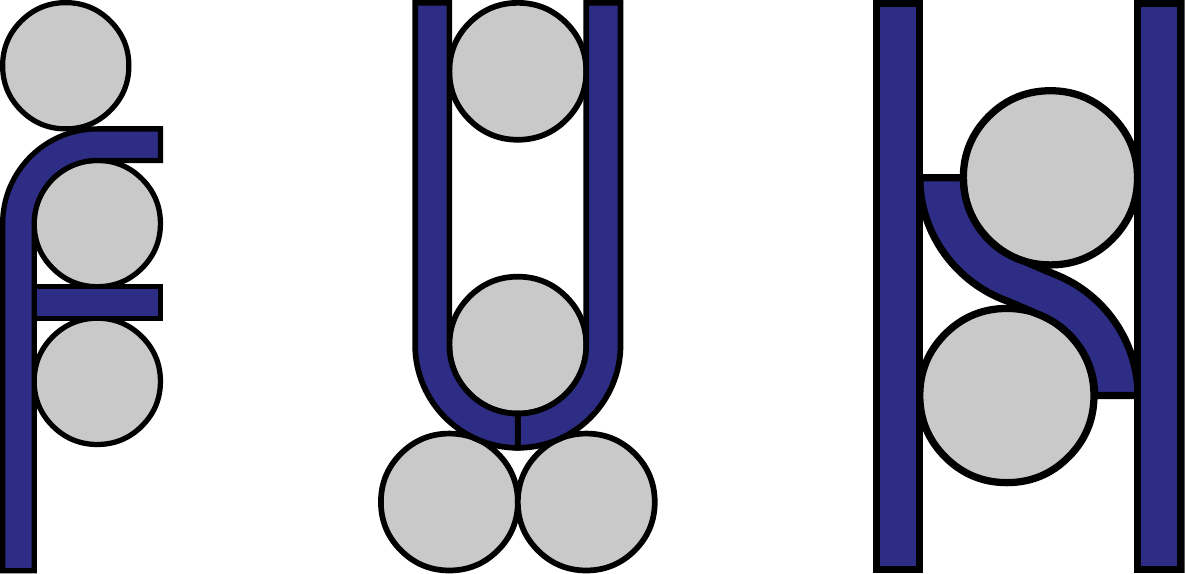}\hfill
  \includegraphics[width=0.5\linewidth]{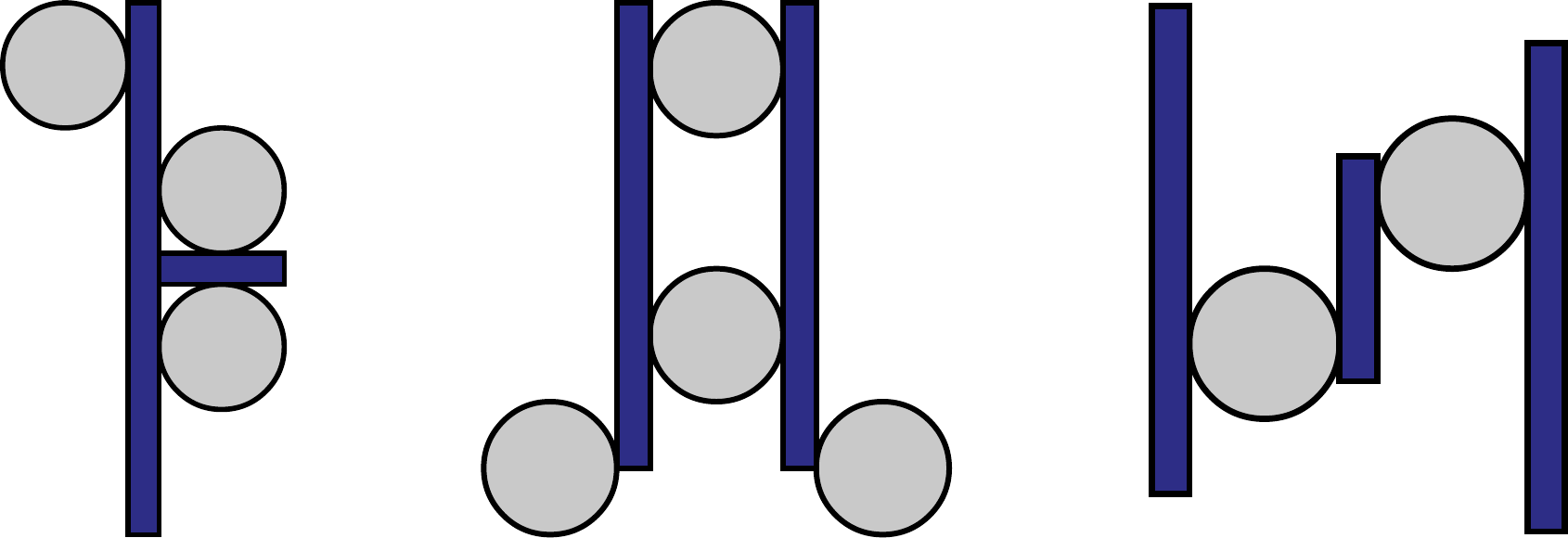}

  \medskip
  \includegraphics[width=0.4\linewidth]{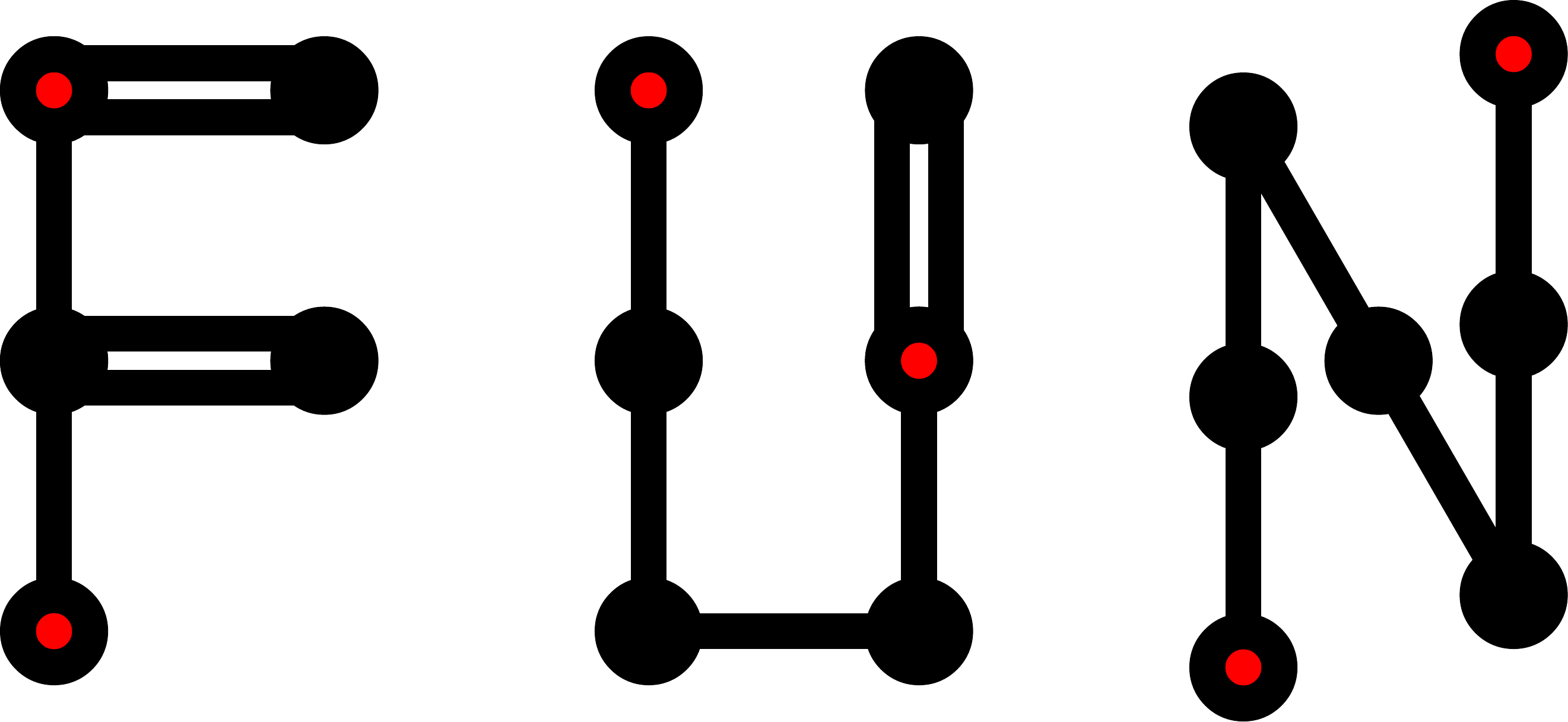}\hfill
  \includegraphics[width=0.5\linewidth]{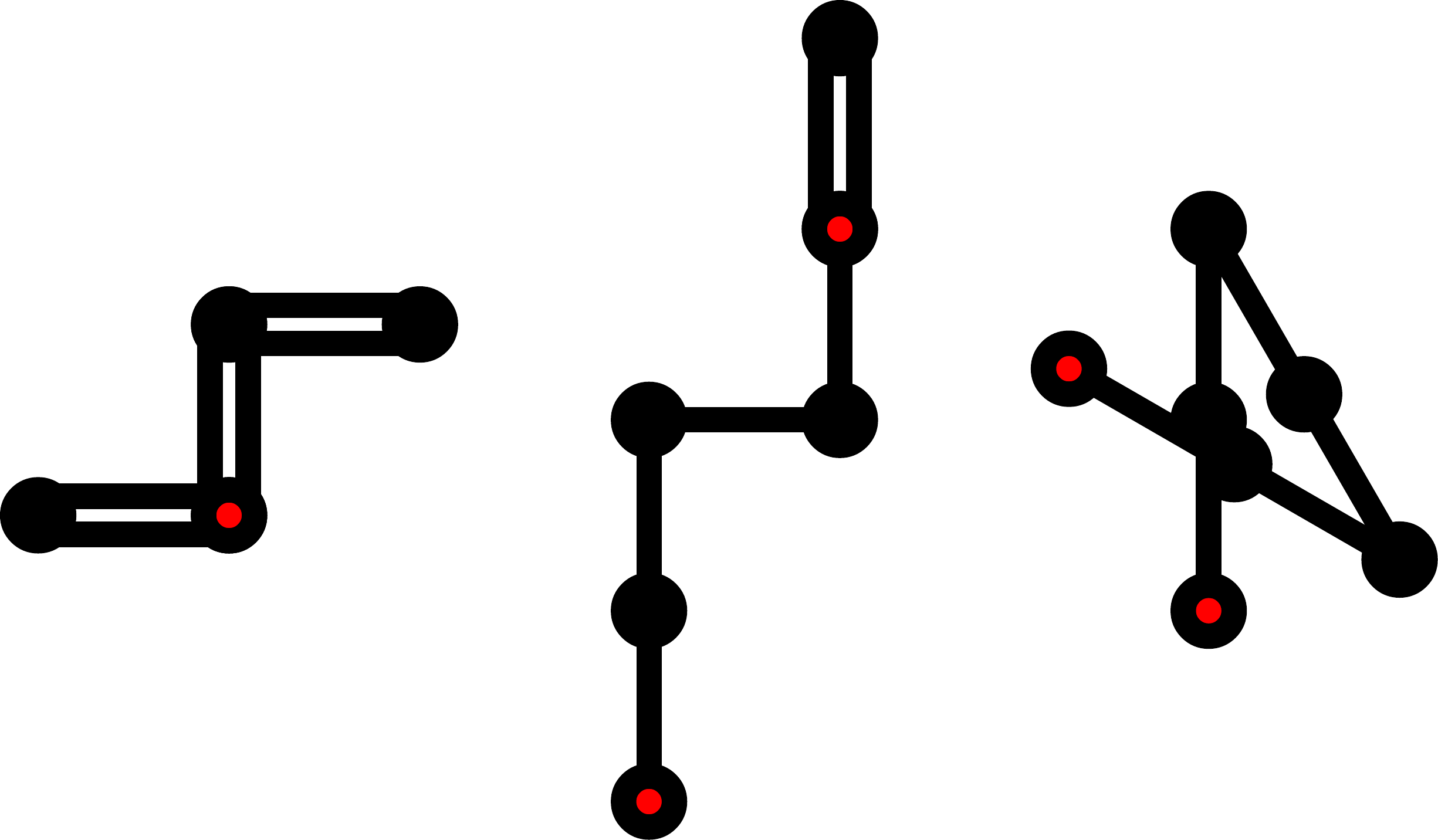}

\iffalse
  \subfloat%[Hinged-dissection typeface]
    {\includegraphics[width=0.33\linewidth]{hinged/fun}}

  \subfloat%[Conveyer typeface, solved with belt]
    {\includegraphics[width=0.45\linewidth]{conveyer/fun}}\hfill
  \subfloat%[Conveyer typeface, puzzle without belt]
    {\includegraphics[width=0.45\linewidth]{conveyer/fun_puzzle}}

  \subfloat%[Origami-maze typeface, 3D extrusion]
    {\includegraphics[width=0.3\linewidth]{maze/fun_3D}}\hfill
  \subfloat%[Origami-maze typeface, puzzle crease pattern]
    {\includegraphics[width=0.6\linewidth]{maze/fun_cp}}

  \subfloat%[Glass-cane typeface]
    {\begin{tabular}{cc}
       \includegraphics[height=0.3in]{cane/top/F} &
       \includegraphics[angle=90,height=0.3in]{cane/side/F} \\
       \includegraphics[height=0.3in]{cane/top/U} &
       \includegraphics[angle=90,height=0.3in]{cane/side/U} \\
       \includegraphics[height=0.3in]{cane/top/N} &
       \includegraphics[angle=90,height=0.3in]{cane/side/N}
     \end{tabular}}

  \vspace*{-1ex}

  \subfloat%[Glass-squishing typeface, line art after squish]
    {\includegraphics[width=0.4\linewidth]{squish/fun_after}}\hfill
  \subfloat%[Glass-squishing typeface, puzzle line art before squish]
    {\includegraphics[width=0.5\linewidth]{squish/fun_before}}

  \vspace*{-1ex}

  \subfloat%[Linkage typeface, correct font]
    {\includegraphics[width=0.4\linewidth]{linkage/fun}}\hfill
  \subfloat%[Linkage typeface, a puzzle font]
    {\includegraphics[width=0.5\linewidth]{linkage/fun_puzzle}}
\fi

  \caption{FUN written in all six of our mathematical typefaces
    (easily readable fonts on the left, puzzle fonts on the right).}
  \vspace*{-4ex}
  \label{all}
\end{figure}

This survey reviews the six typefaces we have designed so far,
in roughly chronological order.  We describe each specific typeface design
along with the underlying algorithmic field.
Figure~\ref{all} shows the example of ``FUN'' written in all six typefaces.
Anyone can experiment with writing text (and puzzles)
in these typefaces using our free web applications.%
\footnote{\url{http://erikdemaine.org/fonts/}}

\section{Hinged Dissections}

A \emph{hinged dissection} is a hinged chain of blocks that can fold into
multiple shapes.  Although hinged dissections date back over 100 years
\cite{Frederickson-1997}, it was only very recently that we proved that
hinged dissections exist, for any set of polygons of equal area
\cite{Abbott-Abel-Charlton-Demaine-Demaine-Kominers-2012}.
That result was the culmination of many years of exploring the problem,
starting with a theorem that any \emph{polyform}---$n$ identical shapes
joined together at corresponding edges---can be folded from one universal
chain of blocks (for each~$n$)
\cite{Demaine-Demaine-Eppstein-Friedman-1999,Demaine-Demaine-Eppstein-Frederickson-Friedman-2005}.

\begin{figure}
  \centering
  \includegraphics[width=\linewidth]{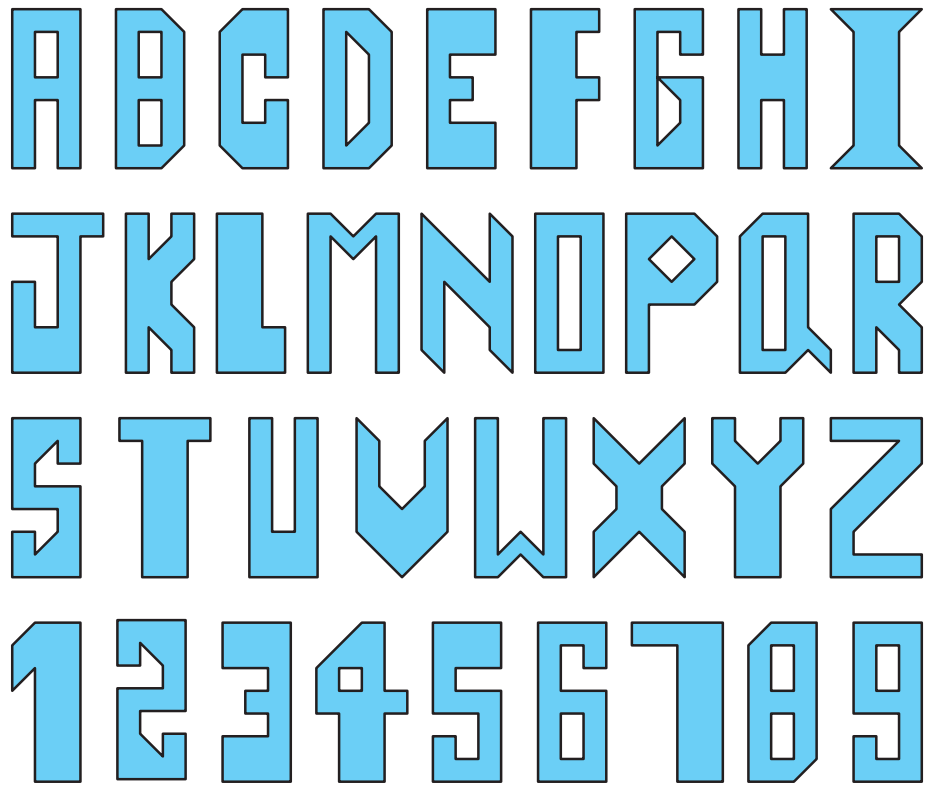}
  \caption{Hinged dissection typeface,
    from \cite{Demaine-Demaine-2003-alphabet}.}
  \label{alphabet outline}
\end{figure}

\begin{figure}
  \centering
  \includegraphics[width=\linewidth]{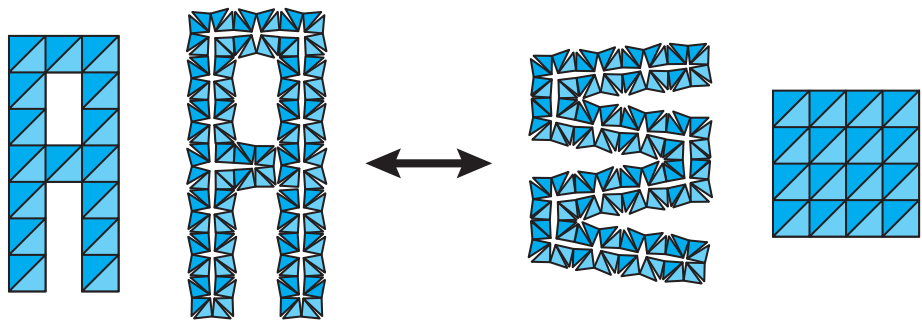}
  \caption{%
    Foldings of the $128$-piece hinged dissection into the letter A and
    a square, from \cite{Demaine-Demaine-2003-alphabet}.}
  \label{a_square}
\end{figure}

Our first mathematical/algorithmic typeface, designed in 2003
\cite{Demaine-Demaine-2003-alphabet},%
\footnote{\url{http://erikdemaine.org/fonts/hinged/}}
illustrates
both this surprising way to hinge-dissect exponentially many polyform shapes,
and the general challenge of the then-open hinged-dissection problem.
As shown in Figure~\ref{alphabet outline}, we designed a series of glyphs
for each letter and numeral as $32$-abolos, that is, edge-to-edge gluings
of 32 identical right isosceles triangles (half unit squares).
In particular, every glyph has the same area.  Applying our theorem about
hinged dissections of polyforms \cite{Demaine-Demaine-Eppstein-Friedman-1999,Demaine-Demaine-Eppstein-Frederickson-Friedman-2005}
produces the $128$-piece hinged dissection shown in Figure~\ref{a_square}.
This universal chain of blocks can fold into any letter in
Figure~\ref{alphabet outline}, as well as a $4 \times 4$ square
as shown in Figure~\ref{a_square}.

An interesting open problem about this font is whether the chain of
$128$ blocks can be folded continuously without self-intersection
into each of the glyphs.
In general, hinged chains of triangles can lock
\cite{Connelly-Demaine-Demaine-Fekete-Langerman-Mitchell-Ribo-Rote-2010}.
But if the simple structure of this hinged dissection enables continuous
motions, we could make a nice animated font, where each letter folds back
and forth between the informationless open chain (or square)
and its folded state as the glyph.
Given a physical instantiation of the chain (probably too large to be
practical), each glyph is effectively a puzzle to see whether it can be
folded continuously without self-intersection.

It would also be interesting to make a puzzle font within this typeface.
Unfolded into a chain, each letter looks the same, as the hinged dissection
is universal.  We could, however, annotate the chain to indicate which parts
touch which parts in the folded state, to uniquely identify each glyph
(after some puzzling).

\section{Conveyer Belts}

A seemingly simple yet still open problem posed by
Manual Abellanas in 2001 \cite{Abellanas-2008} asks whether every disjoint
set of unit disks (gears or wheels) in the plane can be visited by a single taut
non-self-intersecting conveyer belt.  Our research with Bel\'en Palop
first attempted to solve this problem, and then transformed into a new
typeface design \cite{Demaine-Demaine-Palop-2010-elasticity}
and then puzzle design \cite{Demaine-Demaine-Palop-2010-G4G9}.

\begin{figure}
  \centering
  \includegraphics[width=\linewidth]{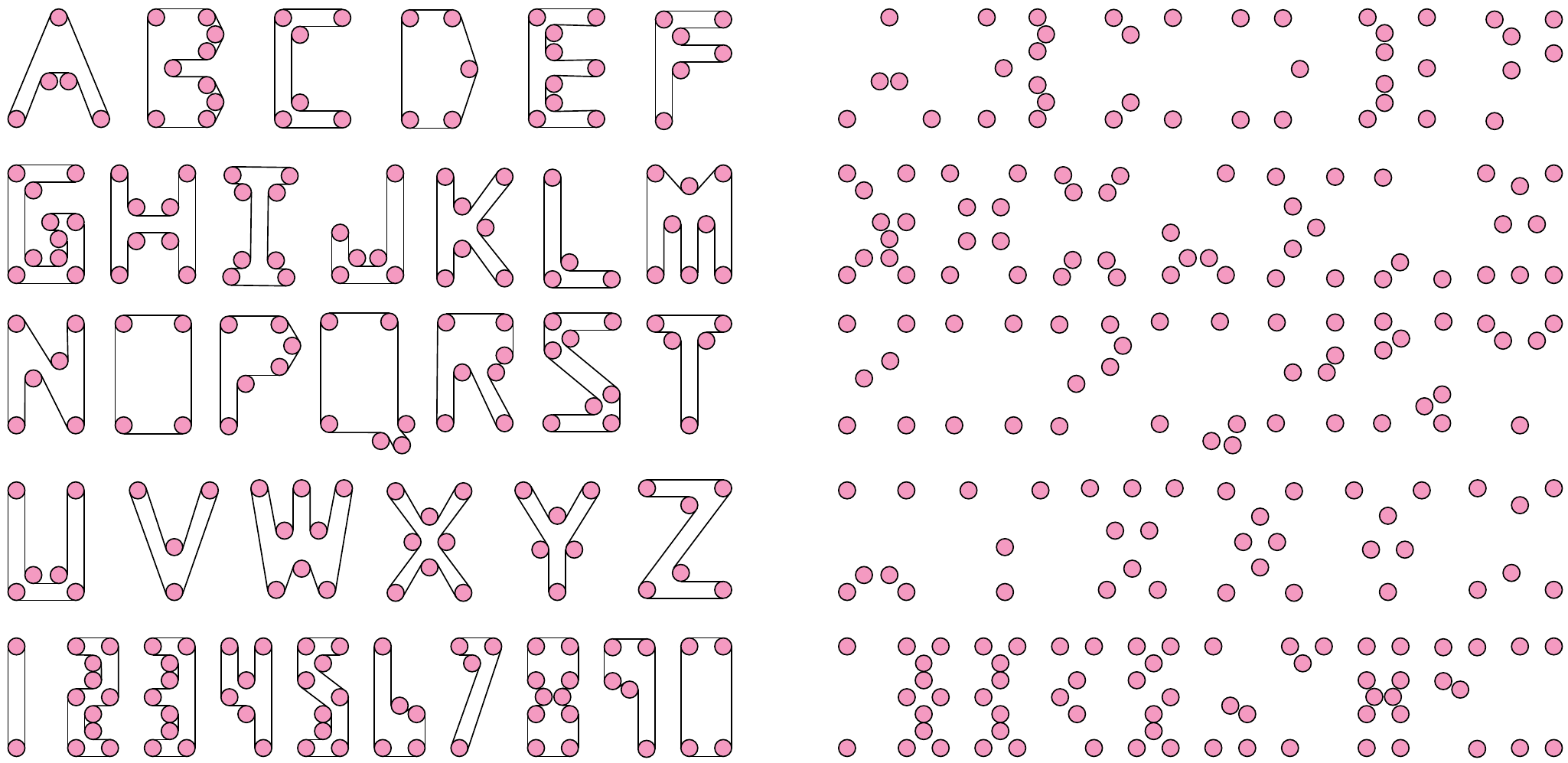}

  \medskip
  \centerline{%
    \hbox to 0.5\linewidth{\hfil (a) With belts.\hfil}\hfil\hfil
    \hbox to 0.5\linewidth{\hfil (b) Without belts.\hfil}%
  }
  \caption{Conveyer belt alphabet,
    from \cite{Demaine-Demaine-Palop-2010-elasticity}.}
  \label{conveyer font}
\end{figure}

The conveyer-belt typeface, shown in Figure~\ref{conveyer font},
consists of all letters and numerals in two main fonts.%
\footnote{\url{http://erikdemaine.org/fonts/conveyer/}}
With both disks and a valid conveyer belt (Figure~\ref{conveyer font}(a)),
the font is easily readable.
But with just the disks (Figure~\ref{conveyer font}(b)),
we obtain a puzzle font where reading each glyph requires solving an instance
of the open problem.  (In fact, each distinct glyph needs to be solved only
once, by recognizing repeated disk configurations.)
Each disk configuration has been designed to have only one solution
conveyer belt that looks like a letter or numeral,
which implies a unique decoding.

\begin{figure}
  \centering
  \includegraphics[width=\linewidth]{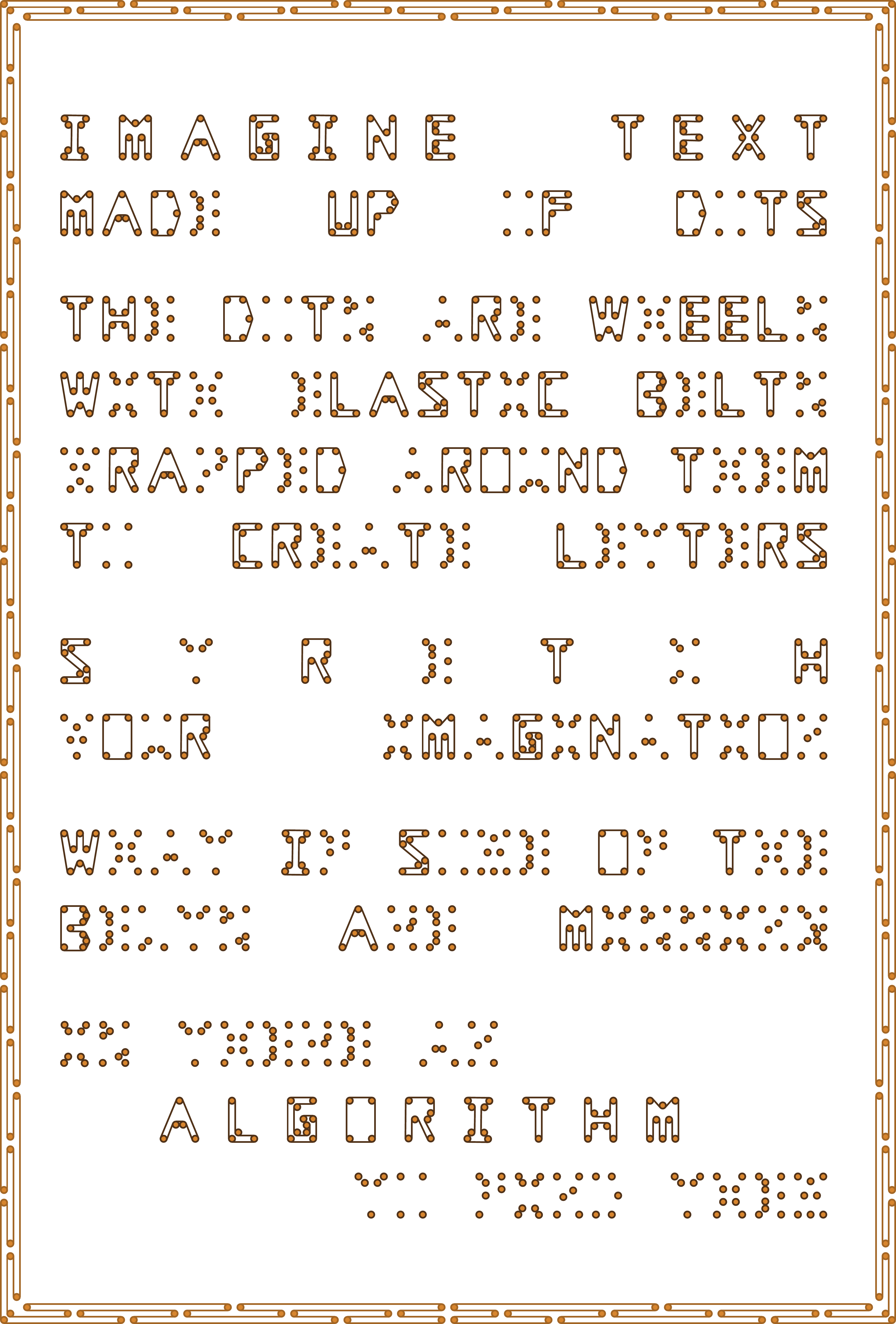}
  \caption{``Imagine Text'' (2013), limited-edition print,
    Erik D. Demaine and Martin L. Demaine, which premiered at the
    Exhibition of Mathematical Art, Joint Mathematics Meetings,
    San Diego, January 2013.}
  %curator: Robert Fathauer
  \label{imagine}
\end{figure}

The puzzle font makes it easy to generate many puzzles with embedded secret 
messages \cite{Demaine-Demaine-Palop-2010-G4G9}.
By combining glyphs from both the puzzle and solved (belt) font,
we have also designed a series of puzzle/art prints.
Figure~\ref{imagine} shows a self-referential puzzle/art print
which describes the very open problem on which it is based.

\section{Origami Mazes}

In computational origami design, the typical goal is to develop algorithms
that fold a desired 3D shape from the smallest possible rectangle of paper
of a desired aspect ratio (typically a square).
One result which achieves a particularly efficient use of paper is
\emph{maze folding} \cite{Demaine-Demaine-Ku-2010}:
any 2D grid graph of horizontal and vertical integer-length segments,
extruded perpendicularly from a rectangle of paper, can be folded from a
rectangle of paper that is a constant factor larger than the target shape.
A striking feature is that the scale factor between the unfolded piece of
paper and the folded shape is independent of the complexity of the maze,
depending only on the ratio of the extrusion height to the maze tunnel width.
(For example, a extrusion/tunnel ratio of $1:1$ induces a
scale factor of $3:1$ for each side of the rectangle.)

\begin{figure}
  \centering
  \subfloat[2D grid maze]{\includegraphics[width=0.5\linewidth]{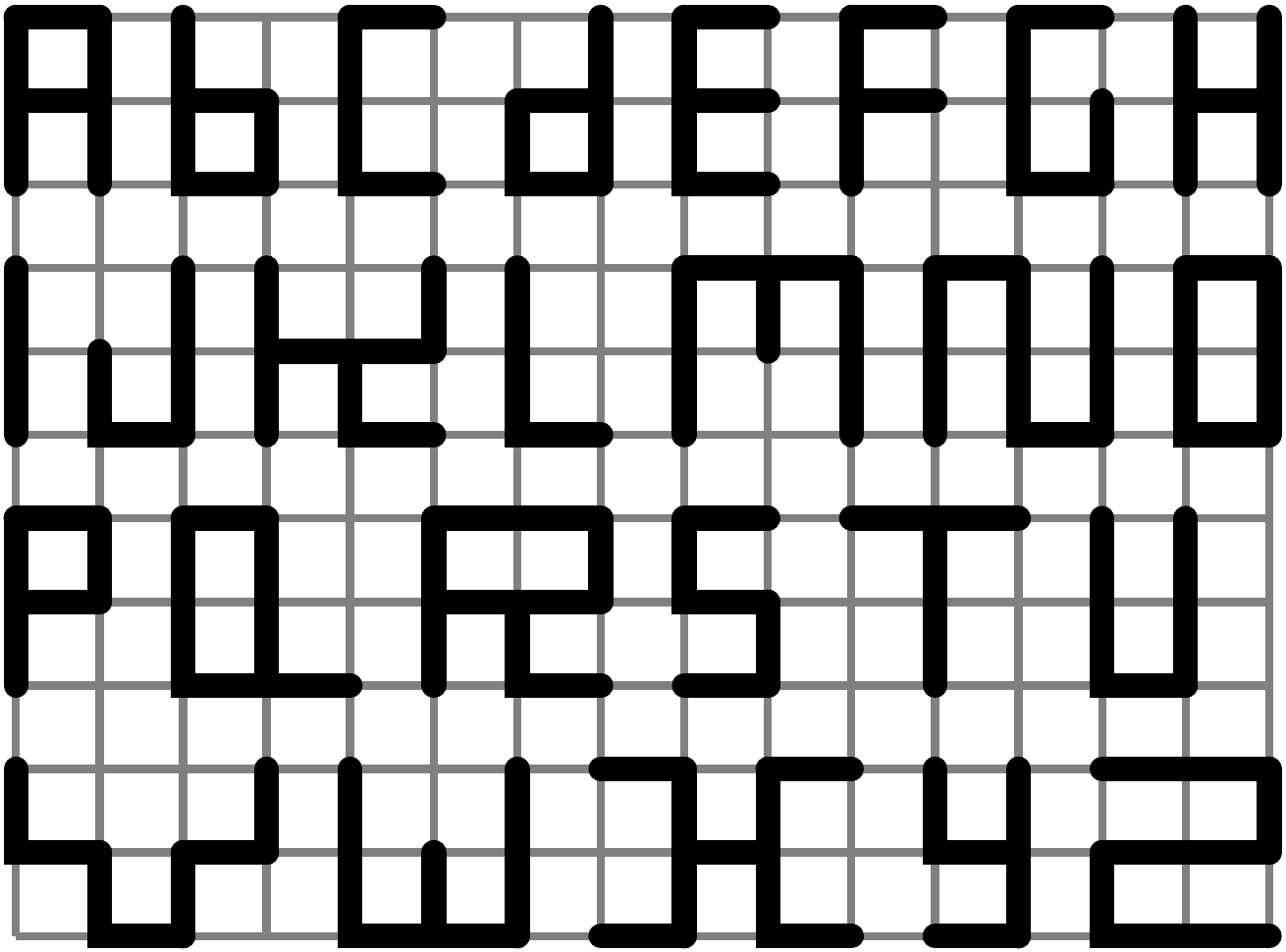}}%
  \hfil\hfil
  \subfloat[3D extrusion]{\includegraphics[width=0.45\linewidth]{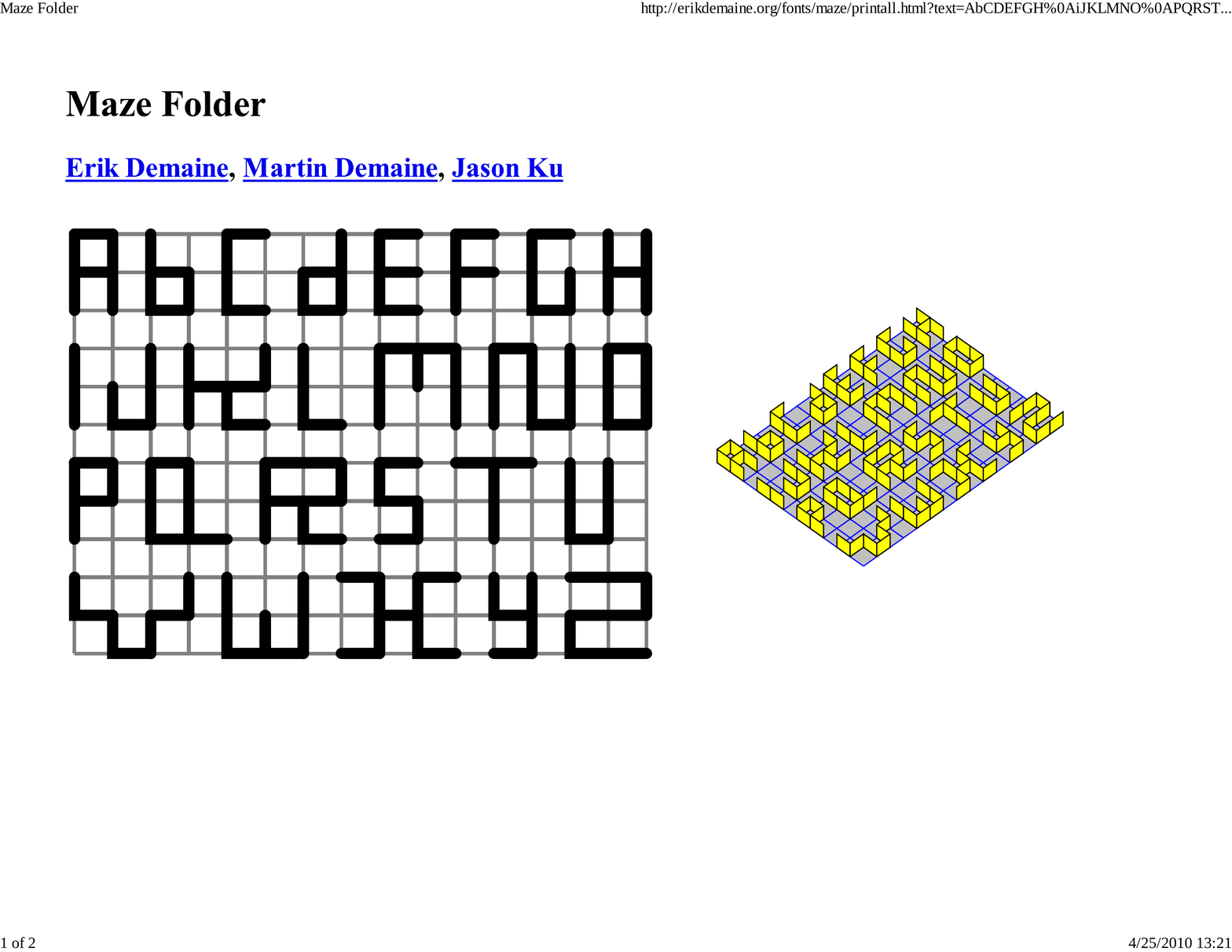}}

  %\subfloat[Crease pattern]{\includegraphics[width=\linewidth]{maze/alphabet_cp}}
  \subfloat[Crease pattern]{\includegraphics[width=\linewidth]{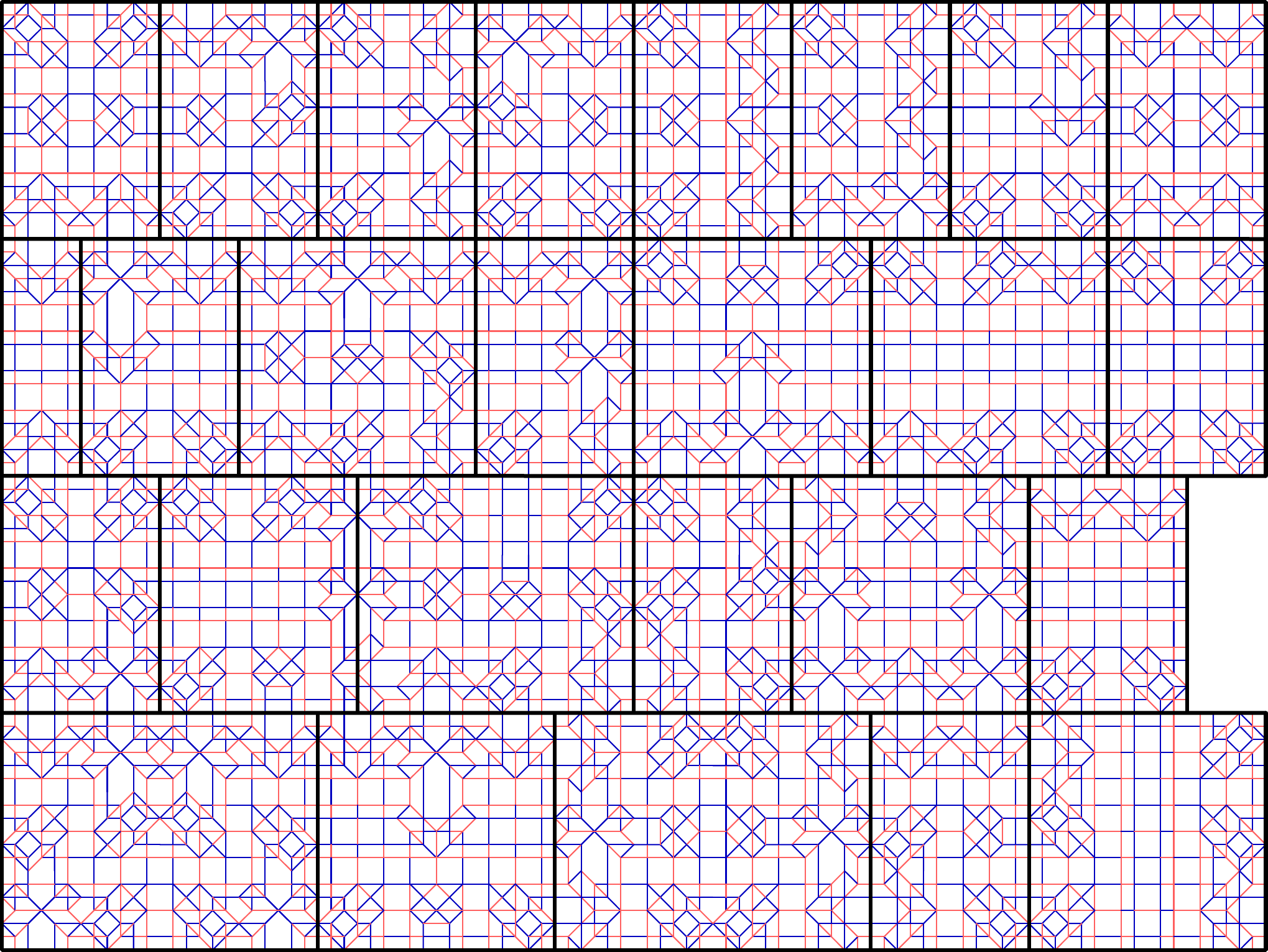}}
  \caption{Origami-maze typeface, from \cite{Demaine-Demaine-Ku-2010-font}:
           (c) folds into (b), which is an extrusion of (a).
           Dark lines are mountain folds; light lines are valley folds;
           bold lines delineate letter boundaries and are not folds.}
  \label{maze font}
  % http://erikdemaine.org/fonts/maze/printall.html?text=AbCDEFGH%0AiJKLMNO%0APQRSTU%0AVWXYZ&lock0=true
  %alternative with numbers:
  % http://erikdemaine.org/fonts/maze/printall.html?text=ABCDEFGHIJ%0AKLMNOPQR%0ASTUVWXYZ%0A1234567890&lock0=true
\end{figure}

The origami-maze typeface, shown in Figure~\ref{maze font},
consists of all letters in three main fonts
\cite{Demaine-Demaine-Ku-2010-font}.%
\footnote{\url{http://erikdemaine.org/fonts/maze/}}
In the 2D font (Figure~\ref{maze font}(a)),
each glyph is written as a 2D grid graph before extrusion.
In the 3D font (Figure~\ref{maze font}(b)),
each glyph is drawn as a 3D extrusion out of a rectangular piece of paper.
In the crease-pattern font (Figure~\ref{maze font}(c)),
each glyph is represented by a crease pattern produced by the maze-folding
algorithm, which folds into the 3D font.
By properties of the algorithm, the crease-pattern font has the feature
that glyphs can be attached together on their boundary to form a larger crease 
pattern that folds into all of the letters as once.
For example, the entire crease pattern of Figure~\ref{maze font}(c)
folds into the 3D shape given by Figure~\ref{maze font}(b).

The crease-pattern font is another puzzle font: each glyph can be read
by folding, either physically or in your head.  With practice, it is possible
to recognize the extruded ridges from the crease pattern alone, and devise
the letters in the hidden message.  We have designed several puzzles along
these lines \cite{Demaine-Demaine-Ku-2010-font}.

\begin{figure}
  \centering
  \includegraphics[width=\linewidth]{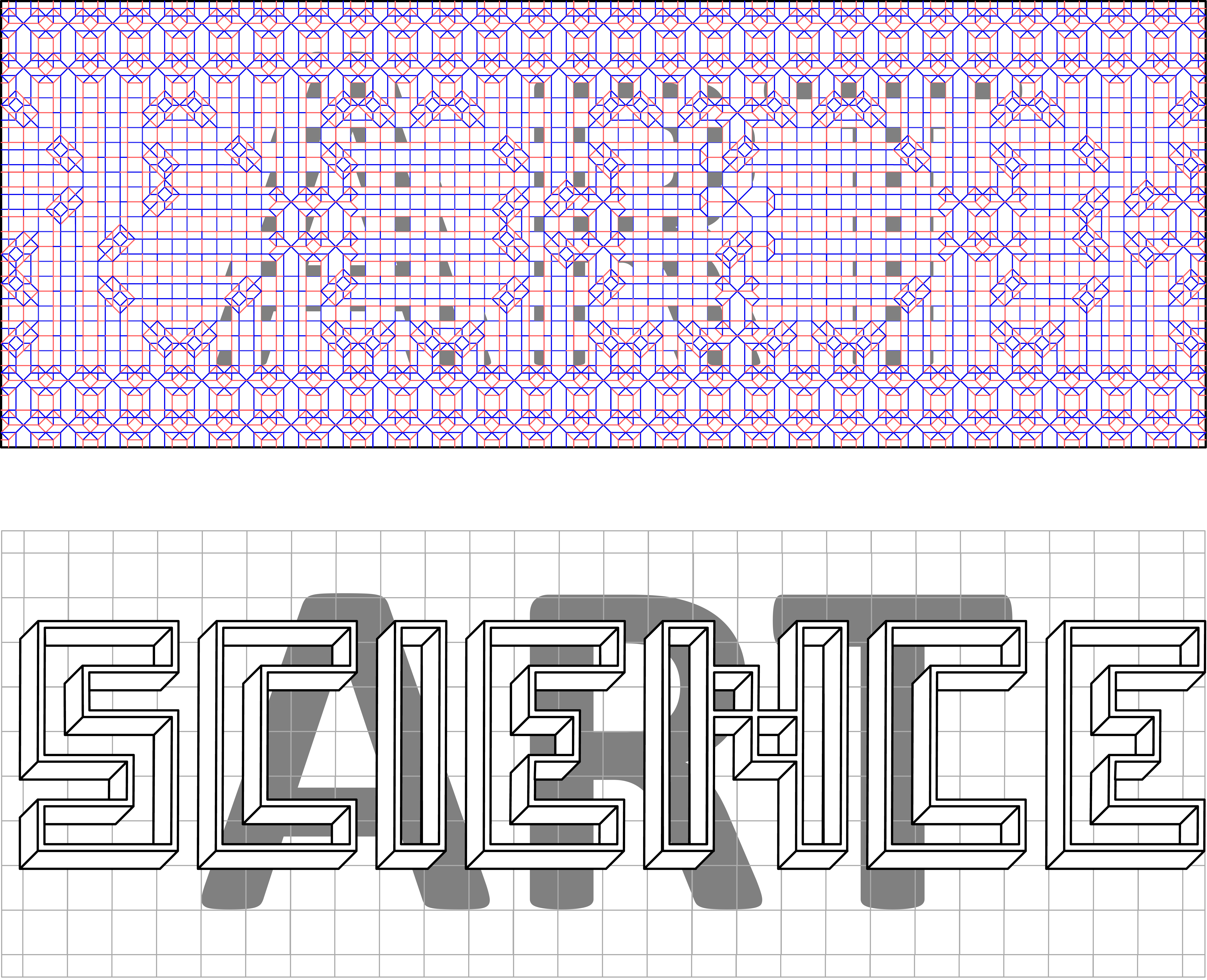}
  \caption{``Science/Art'' (2011), limited-edition print,
    Erik D. Demaine and Martin L. Demaine, which premiered at the
    Exhibition of Mathematical Art, Joint Mathematics Meetings,
    Boston, January 2012.}
  \label{science art}
\end{figure}

It is also possible to overlay a second puzzle within the crease-pattern font,
by placing a message or image in the ground plane of the 3D folded shape,
dividing up by the grid lines, and unfolding those grid cells to where they
belong in the crease pattern.  Figure~\ref{science art} shows one print design
along these lines, with the crease pattern defining the 3D extrusion of
``SCIENCE'' while the gray pattern comes together to spell ``ART''.
In this way, we use our typeface design to inspire new print designs.

\section{Glass Cane}

Glass blowing is an ancient art form, and today it uses most of the same
physical tools as centuries ago.  In computer-aided glass blowing, our goal is
to harness geometric and computational modeling to enable design of glass
sculpture and prediction of how it will look ahead of time on a computer.
This approach enables extensive experimentation with many variations of a
design before committing the time, effort, and expense required to physically
blow the piece.

Our free software Virtual Glass \cite{VirtualGlass-SIGGRAPH2012}
currently focuses on computer-aided design of the highly geometric aspects of
glass blowing, particularly glass cane.  Glass cane is a process for making
long thin cylinders of glass containing elaborate twisty patterns of color.
In Virtual Glass, the user designs the cross-section of a cane by combining
canes and/or colors, and then controls the helical twist of the resulting
form.

\begin{figure}
  \centering
  \tabcolsep=0.5\tabcolsep
  \begin{tabular}{ccccccccccccc}
    \includegraphics[width=0.25in]{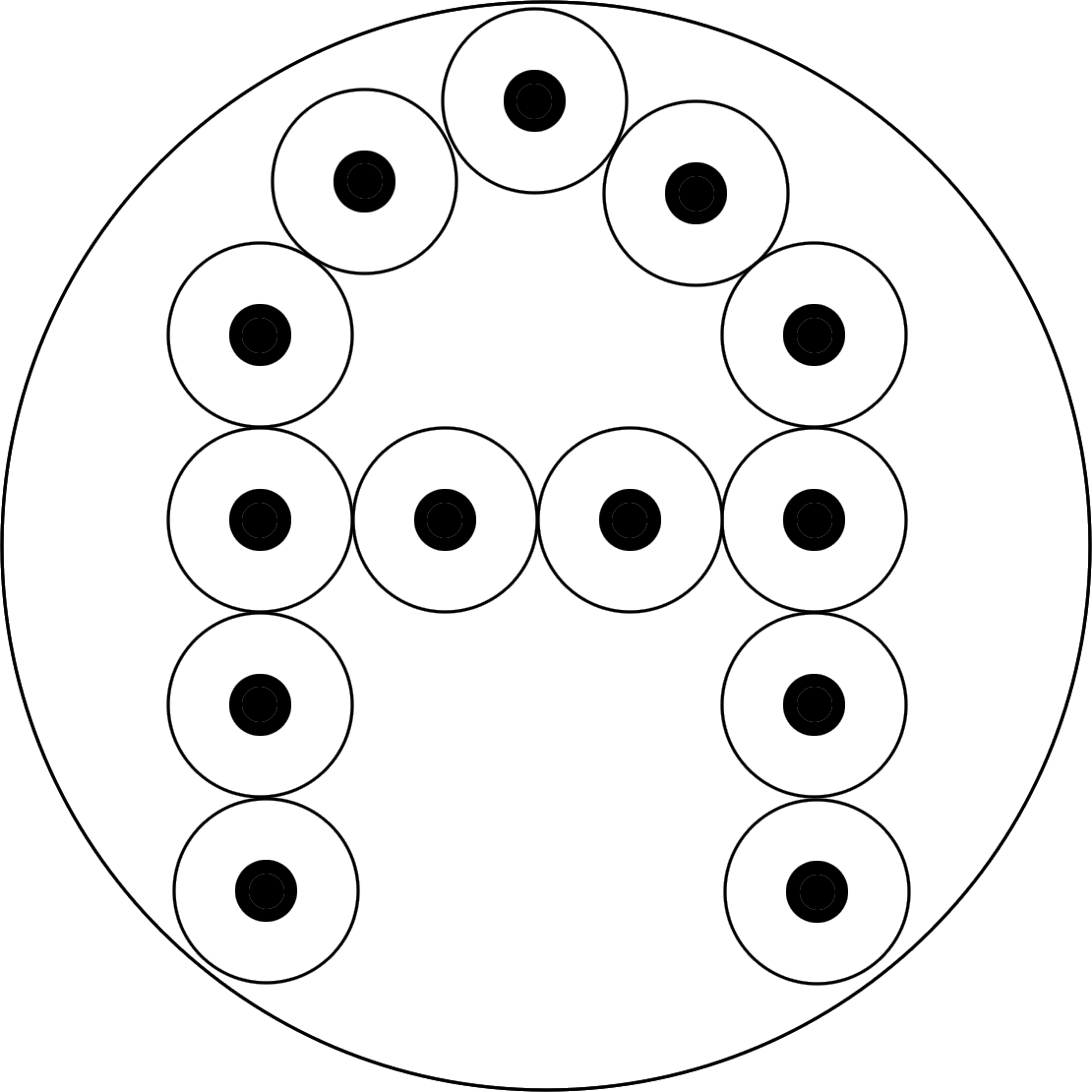} &
    \includegraphics[width=0.25in]{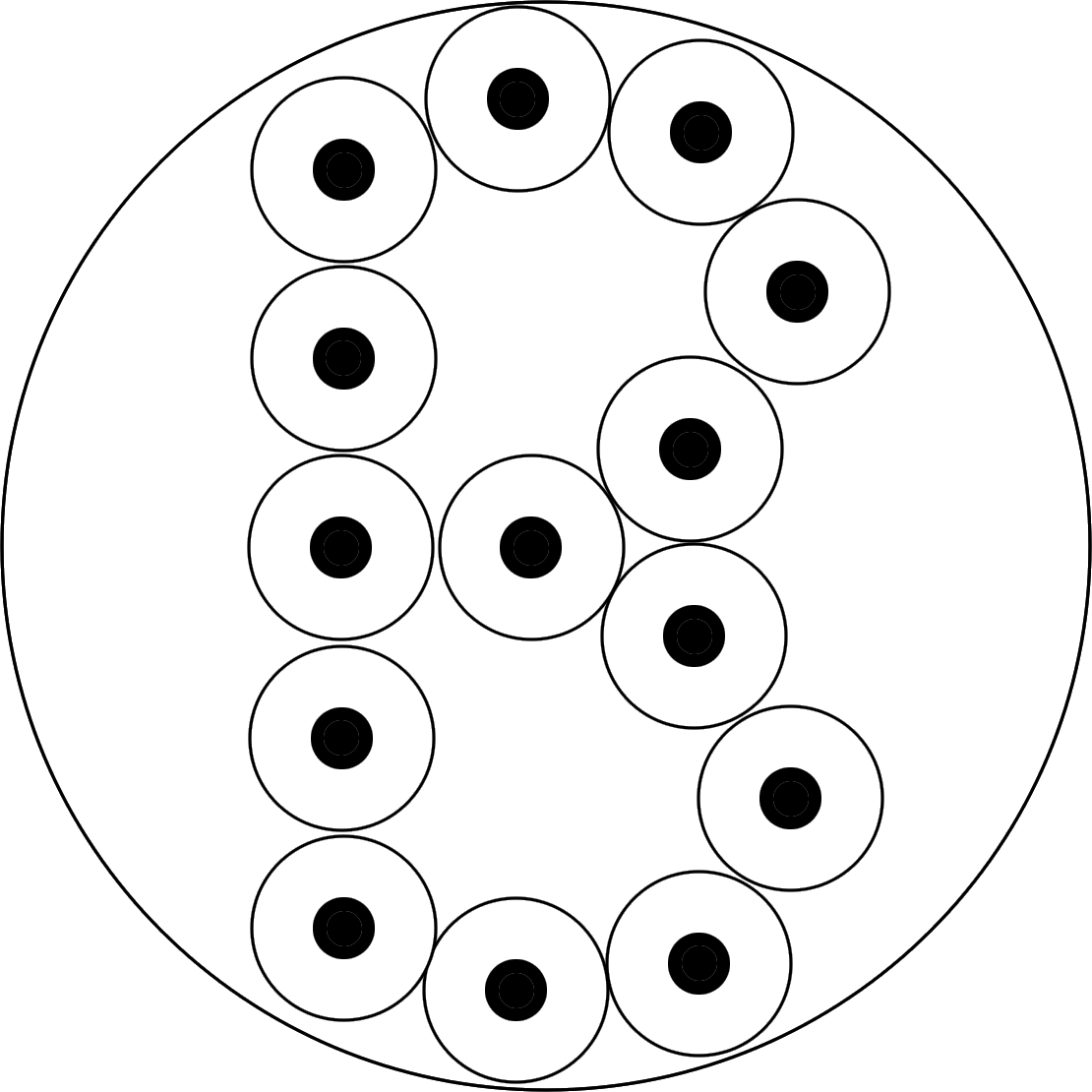} &
    \includegraphics[width=0.25in]{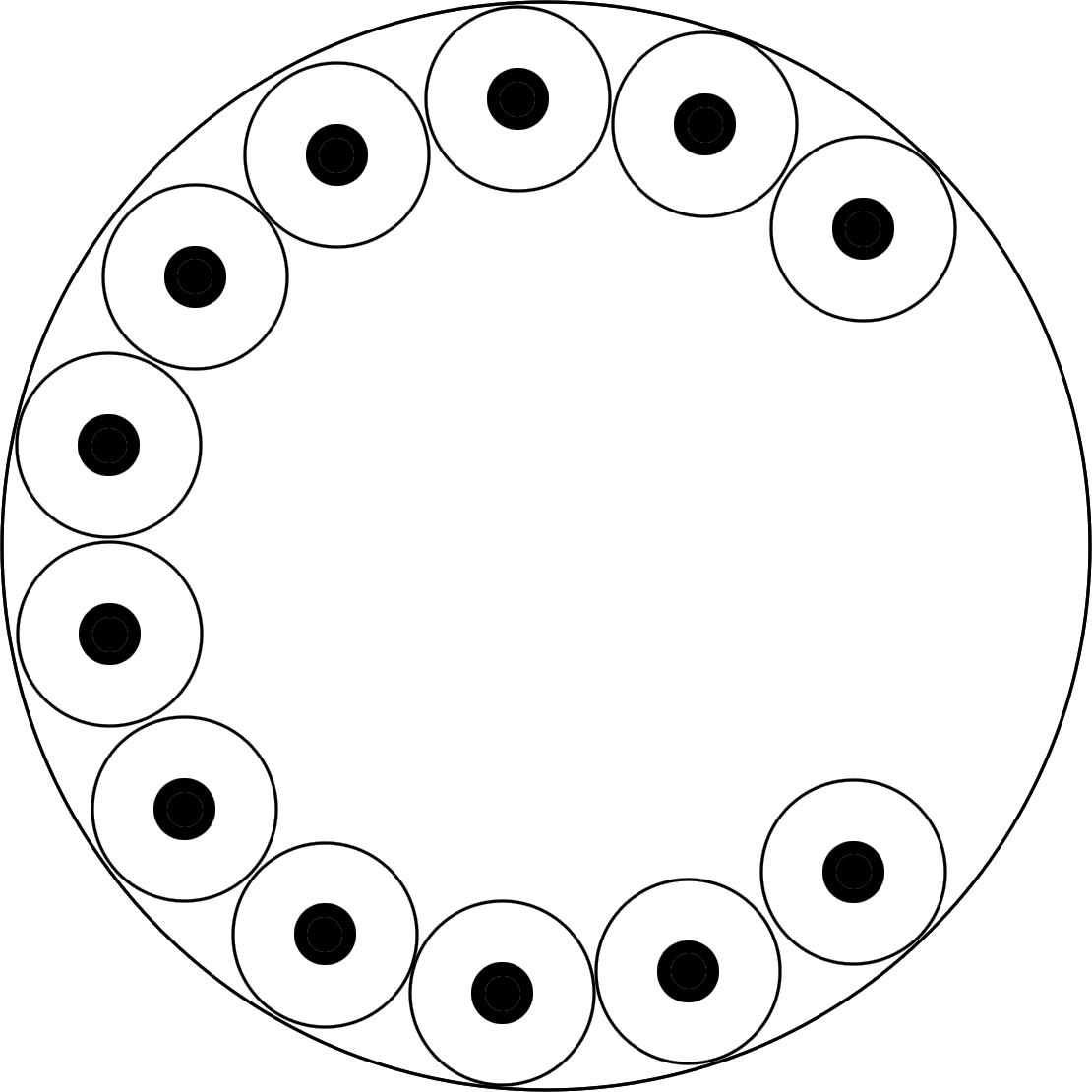} &
    \includegraphics[width=0.25in]{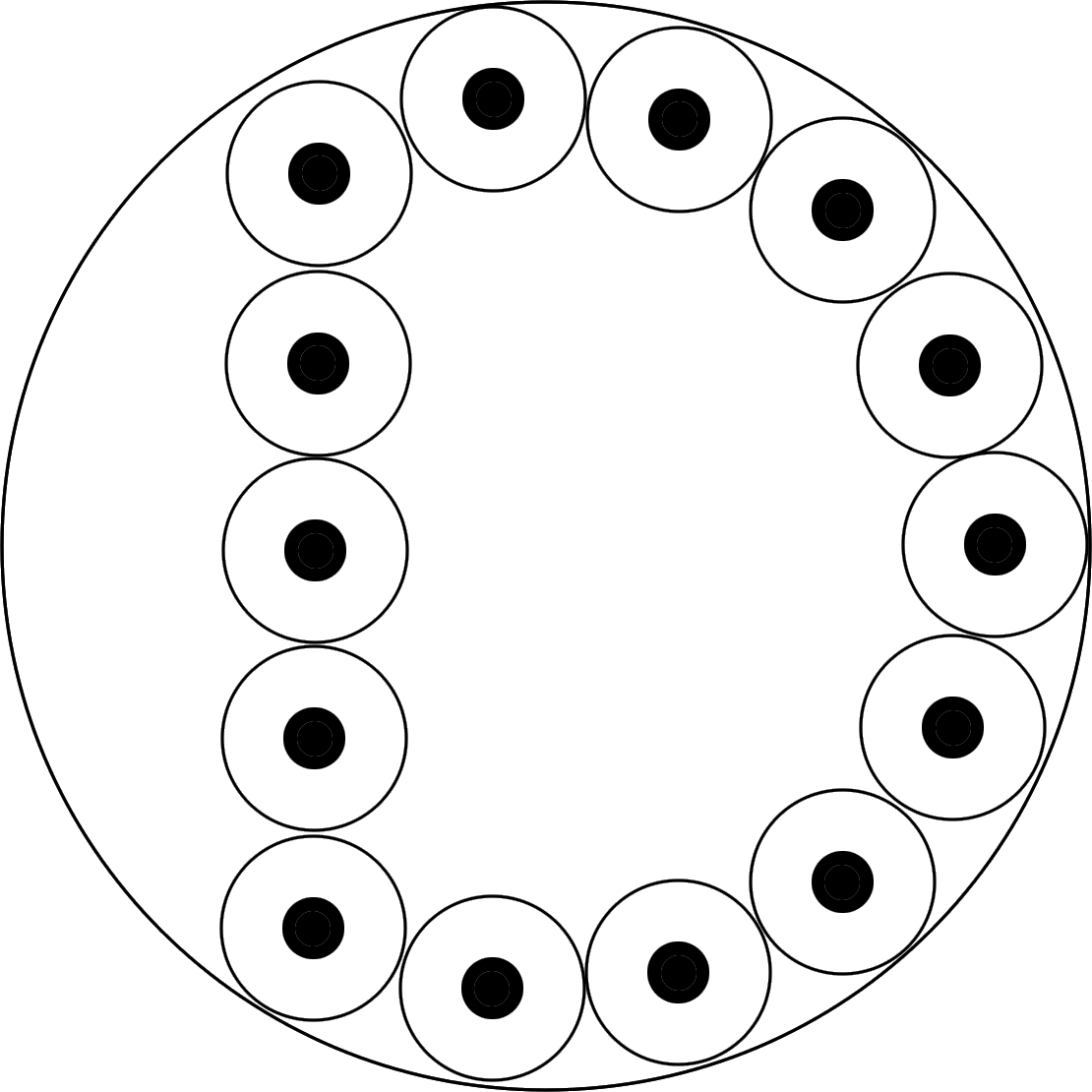} &
    \includegraphics[width=0.25in]{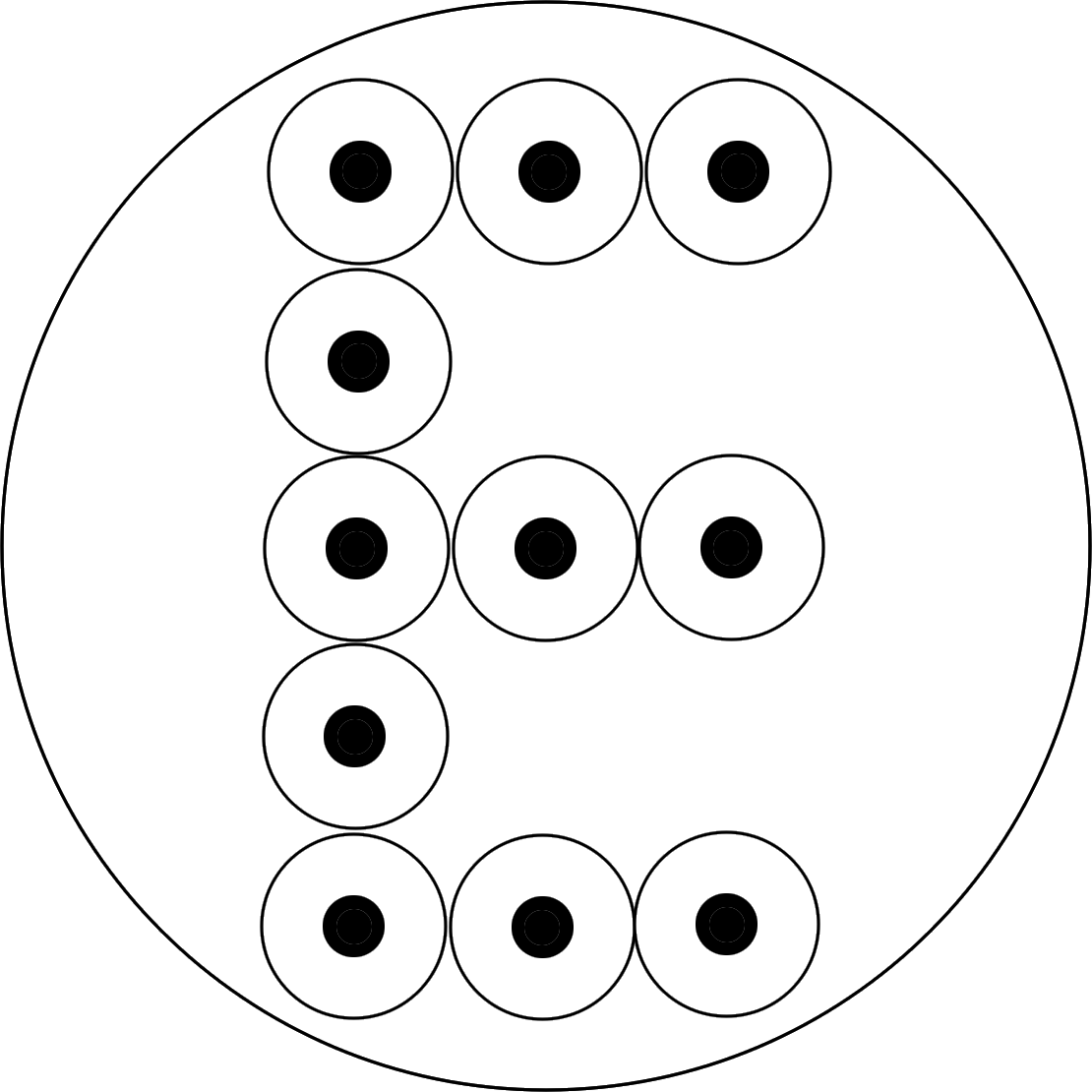} &
    \includegraphics[width=0.25in]{cane/top/F} &
    \includegraphics[width=0.25in]{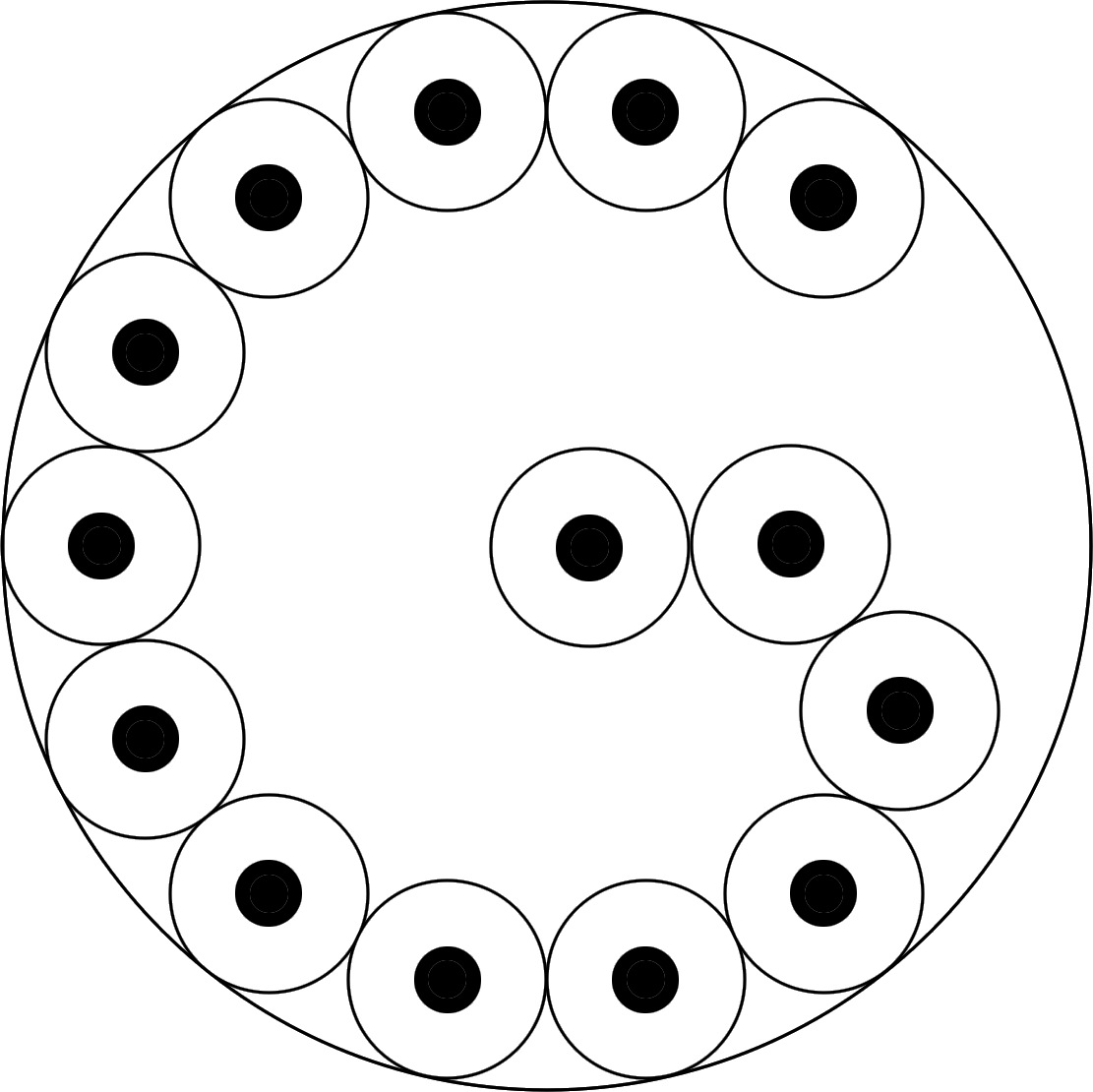} &
    \includegraphics[width=0.25in]{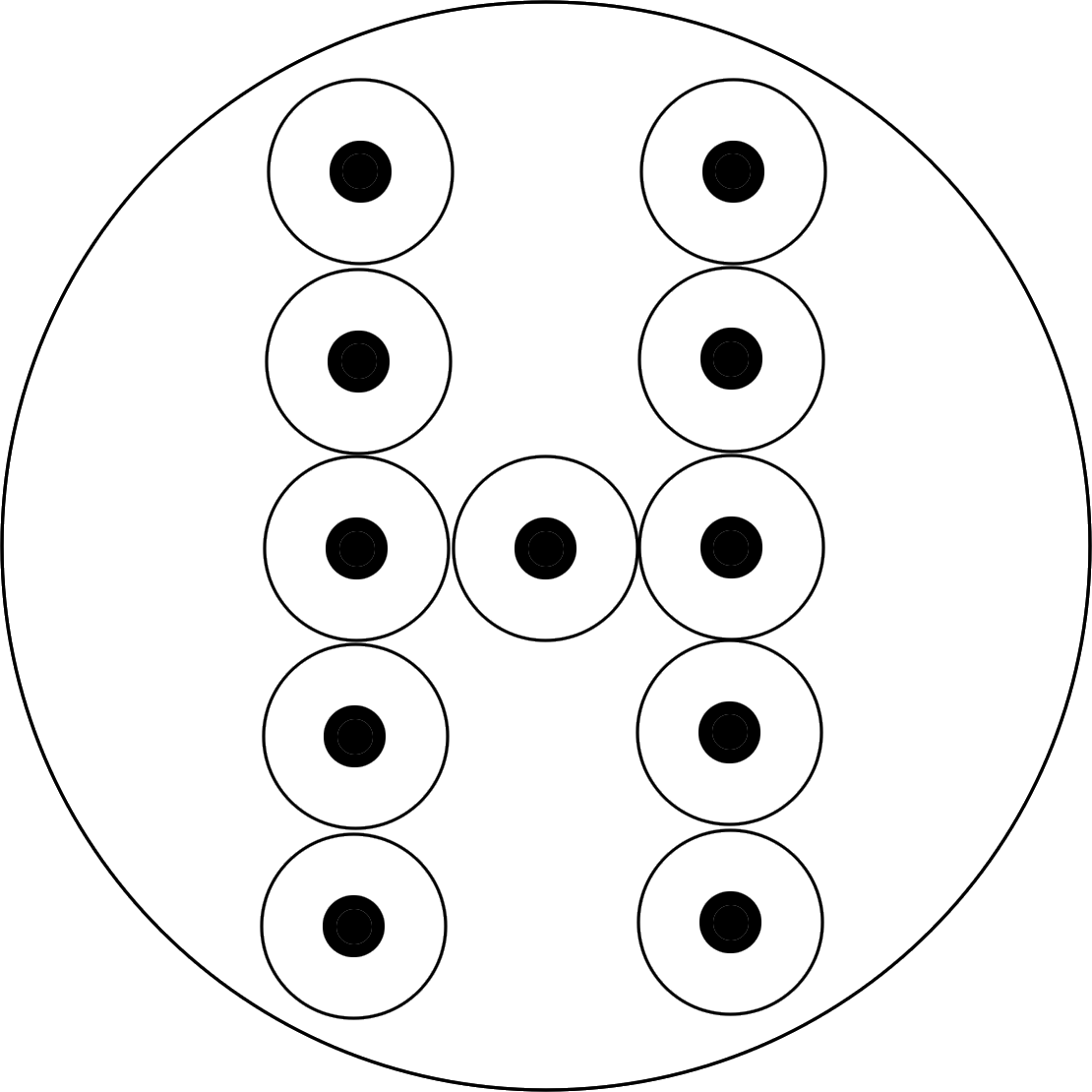} &
    \includegraphics[width=0.25in]{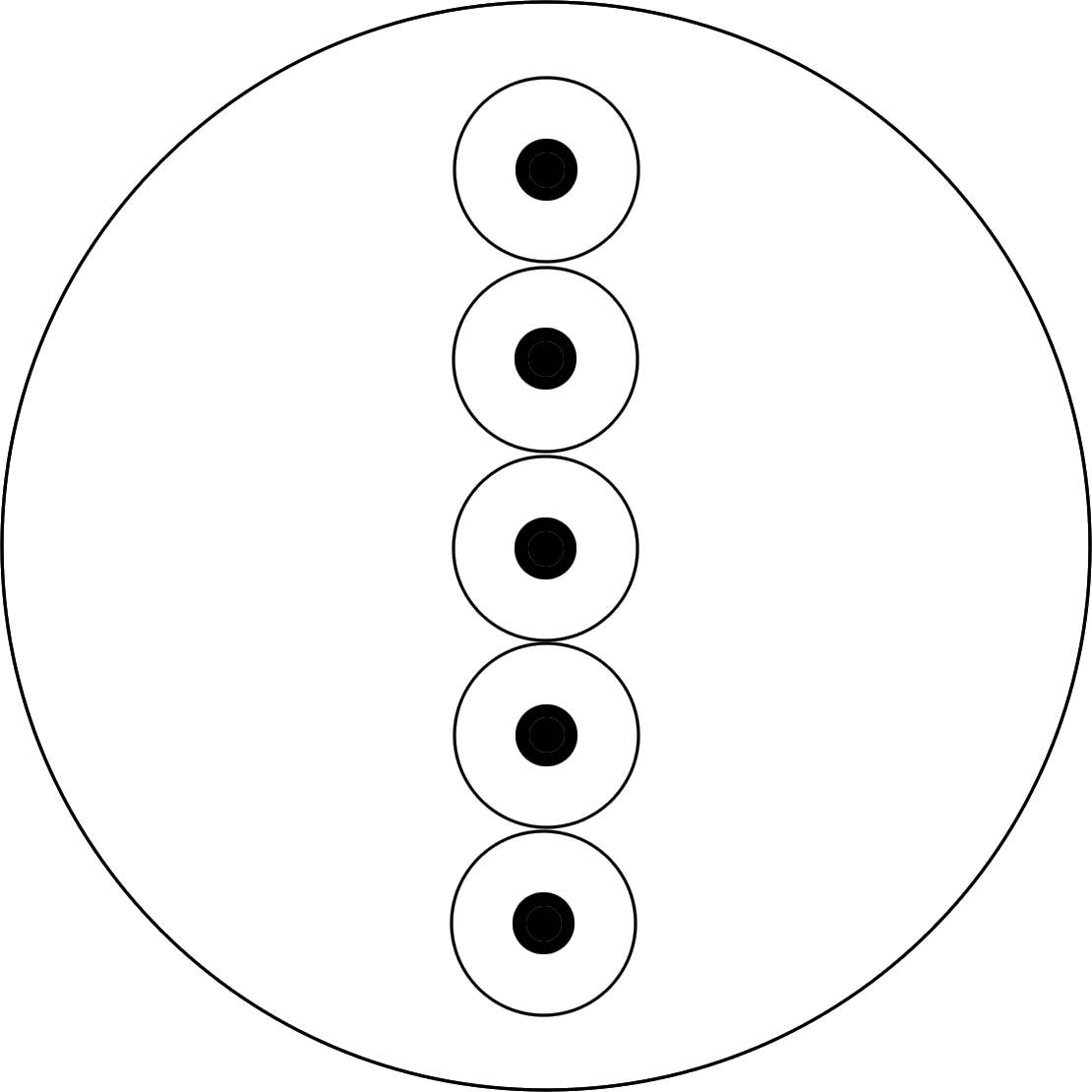} &
    \includegraphics[width=0.25in]{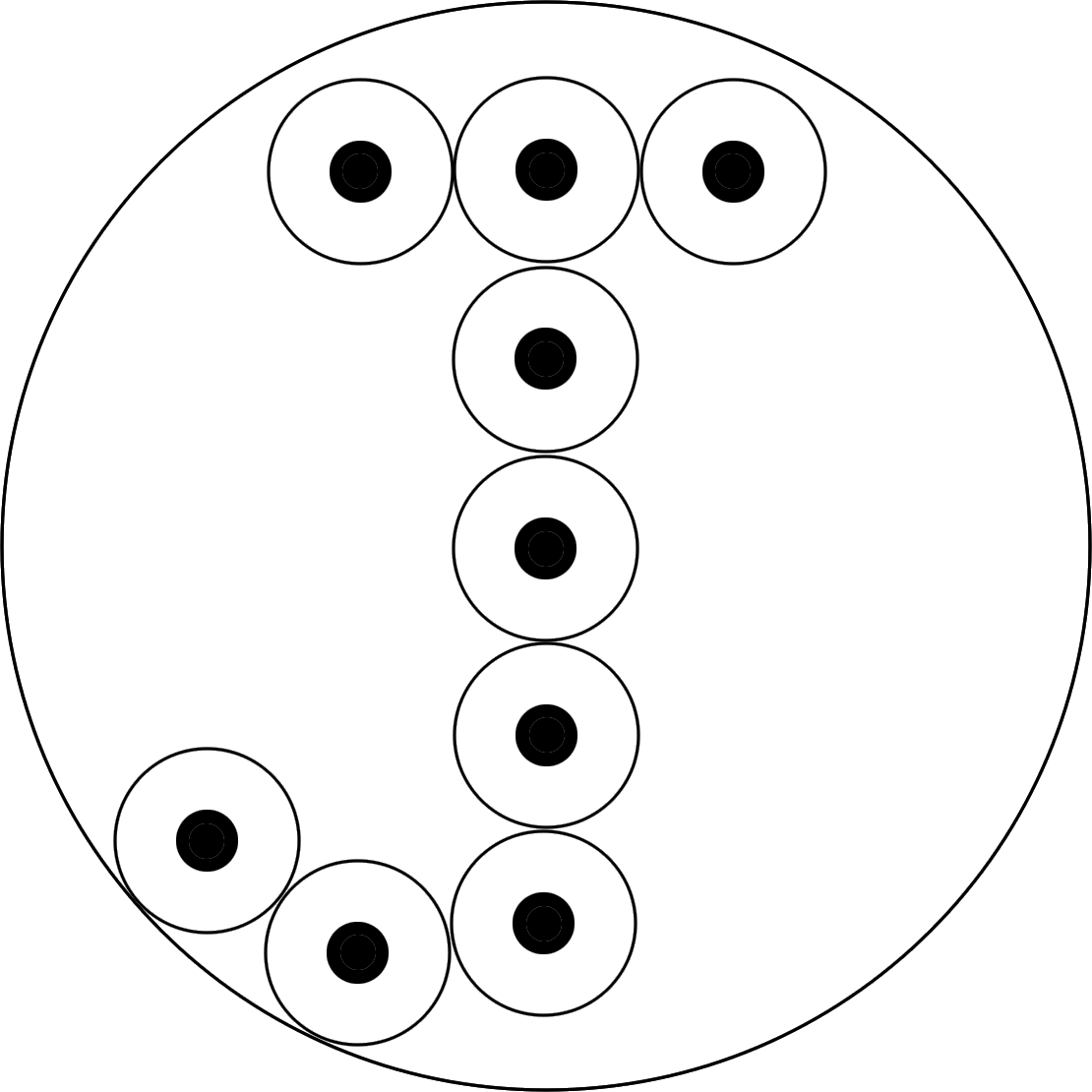} &
    \includegraphics[width=0.25in]{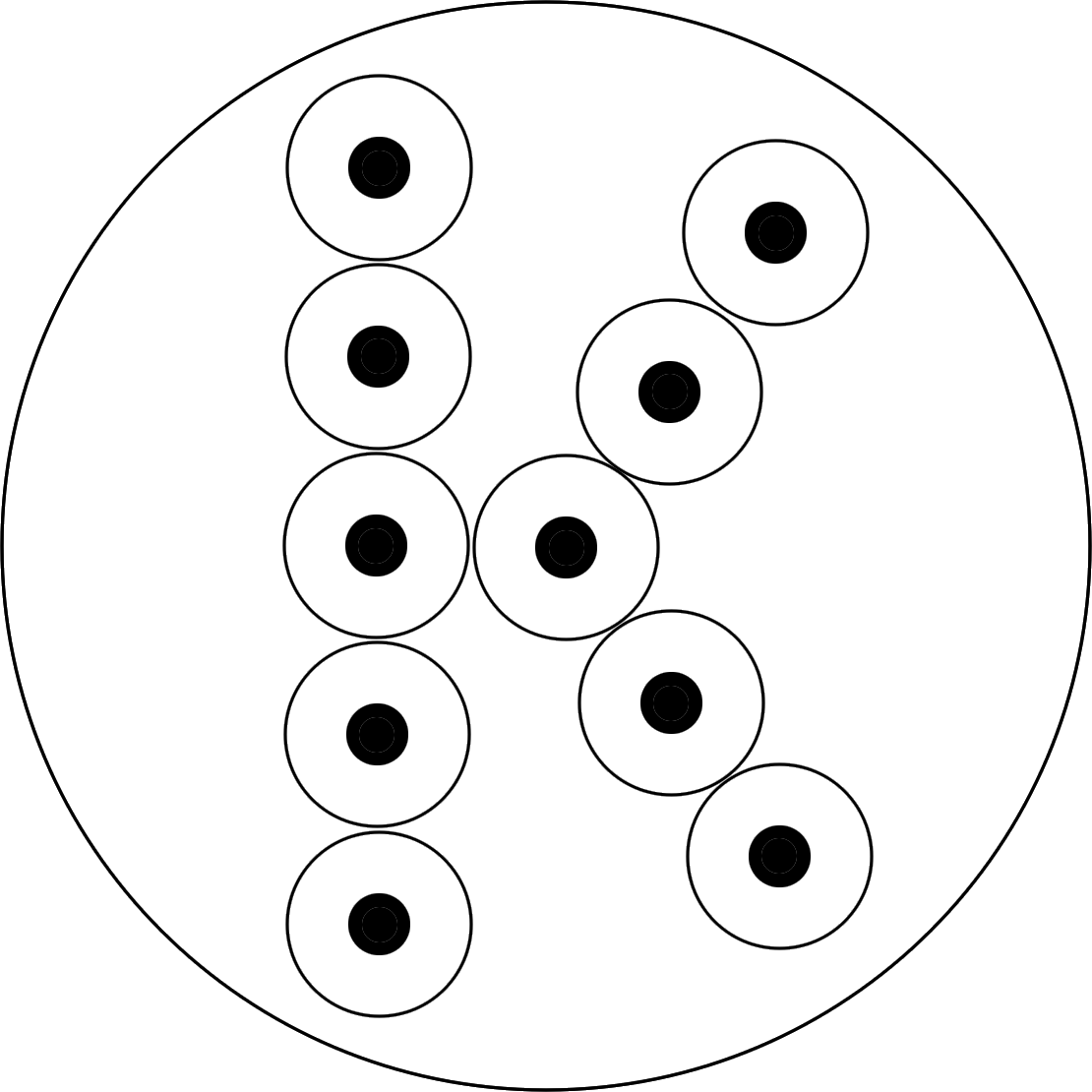} &
    \includegraphics[width=0.25in]{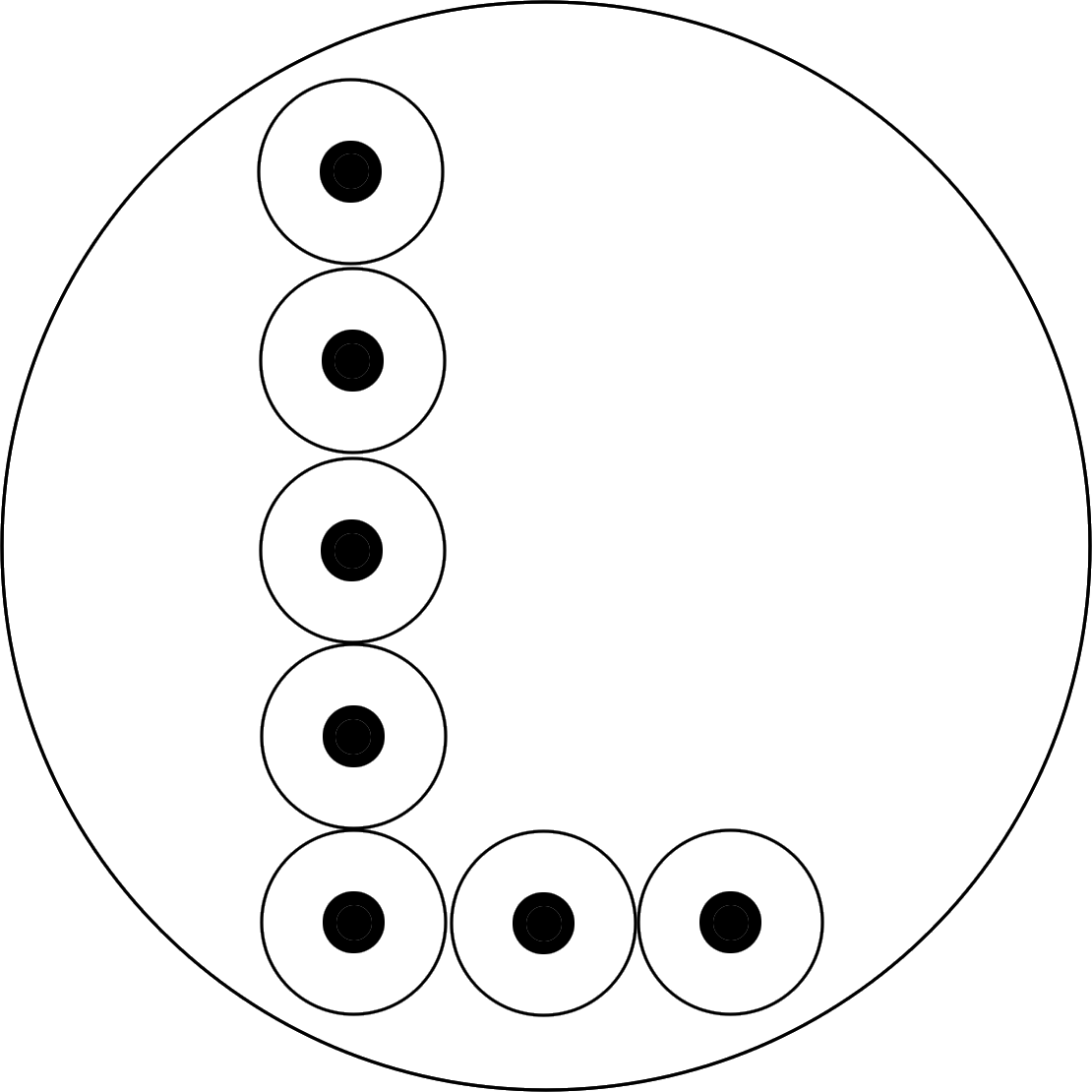} &
    \includegraphics[width=0.25in]{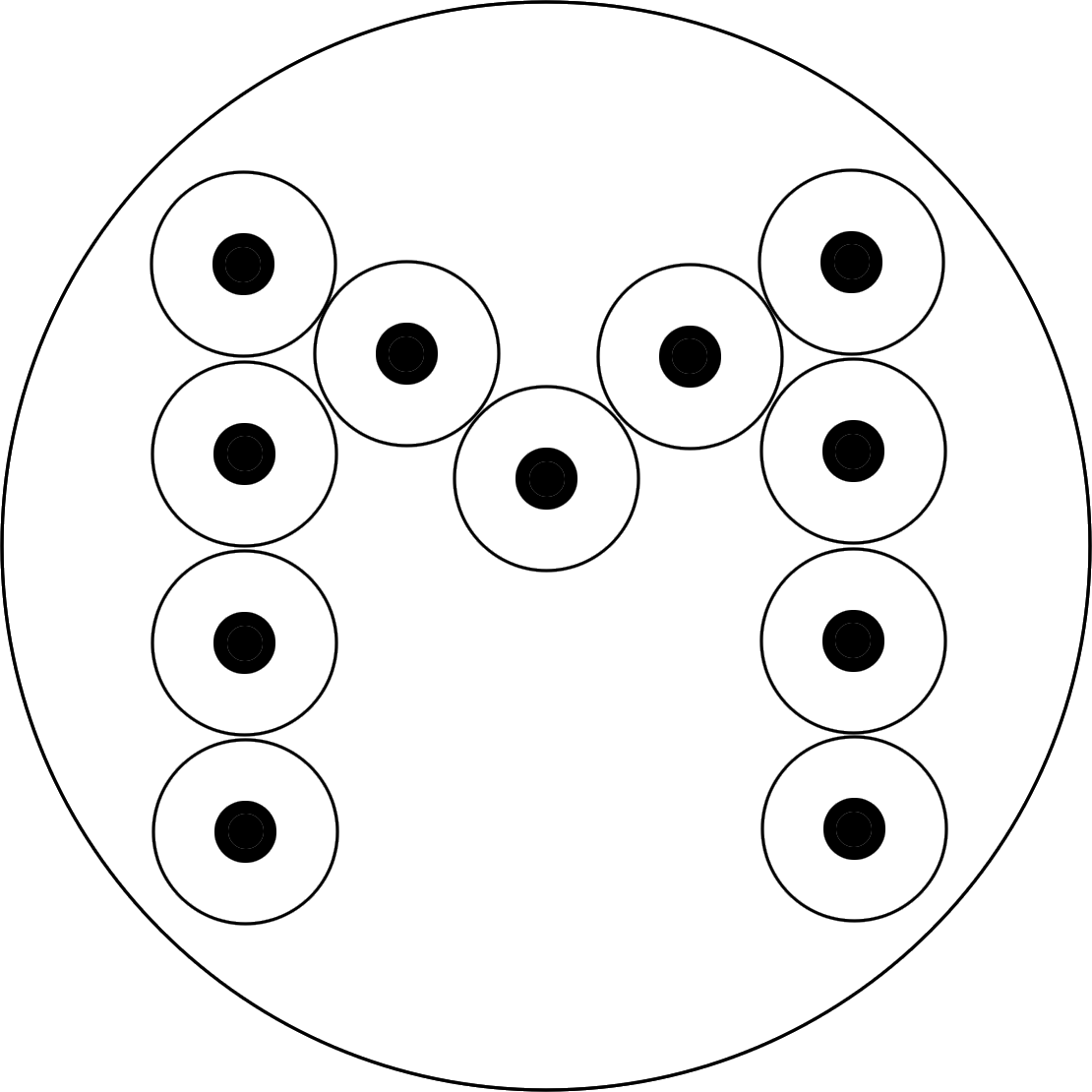}
    \\
    \includegraphics[width=0.25in]{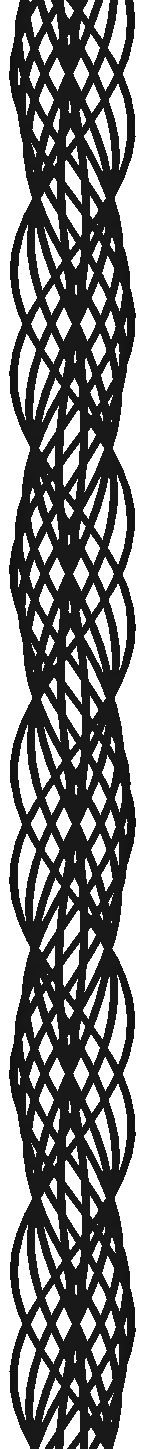} &
    \includegraphics[width=0.25in]{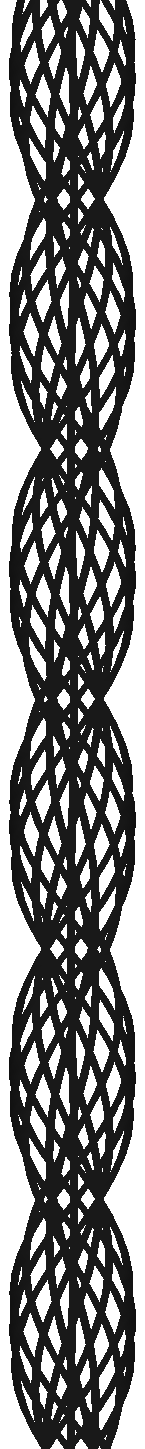} &
    \includegraphics[width=0.25in]{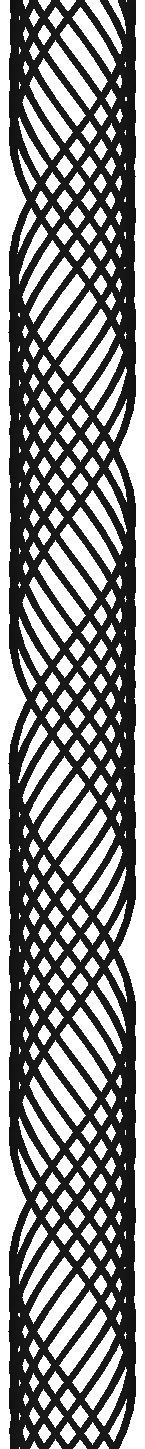} &
    \includegraphics[width=0.25in]{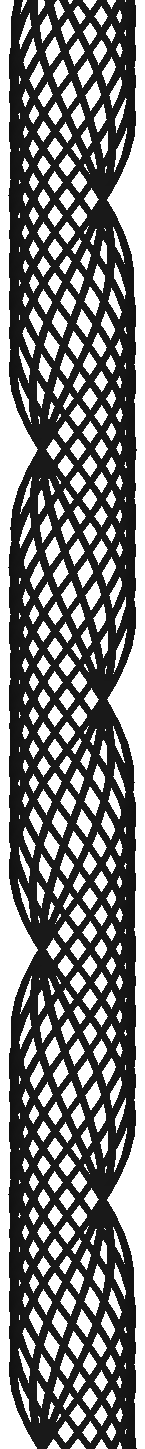} &
    \includegraphics[width=0.25in]{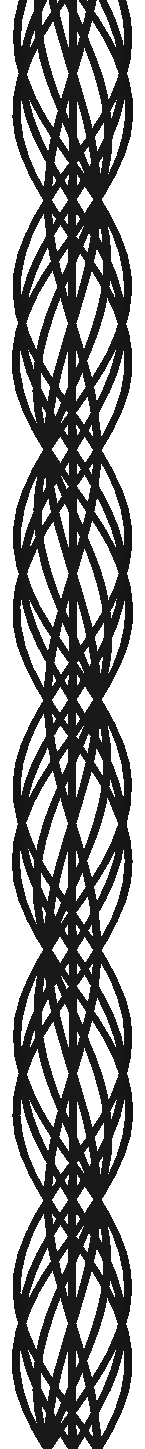} &
    \includegraphics[width=0.25in]{cane/side/F} &
    \includegraphics[width=0.25in]{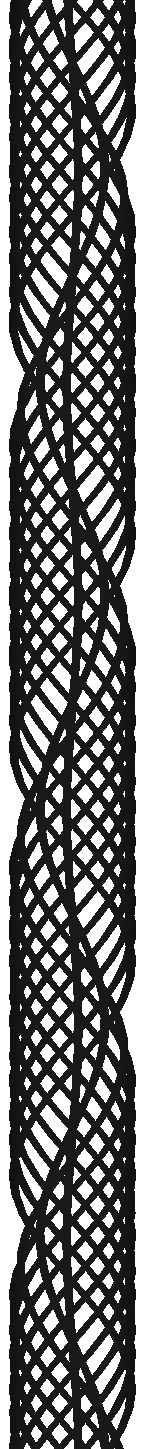} &
    \includegraphics[width=0.25in]{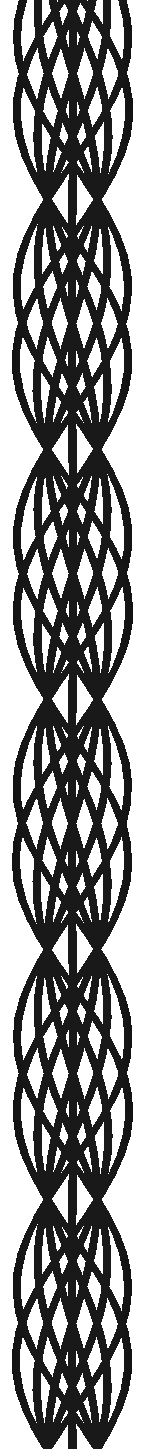} &
    \includegraphics[width=0.25in]{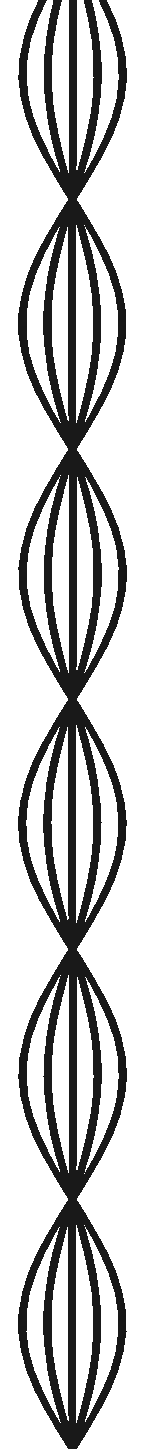} &
    \includegraphics[width=0.25in]{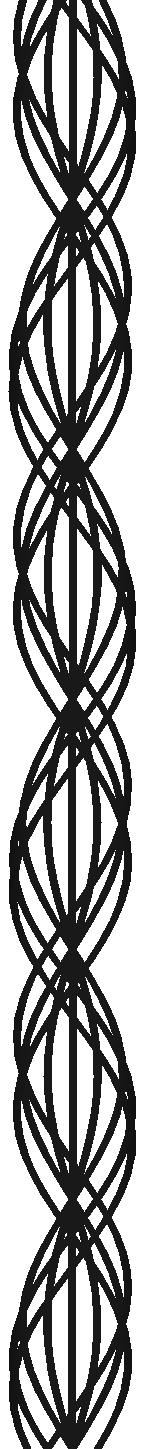} &
    \includegraphics[width=0.25in]{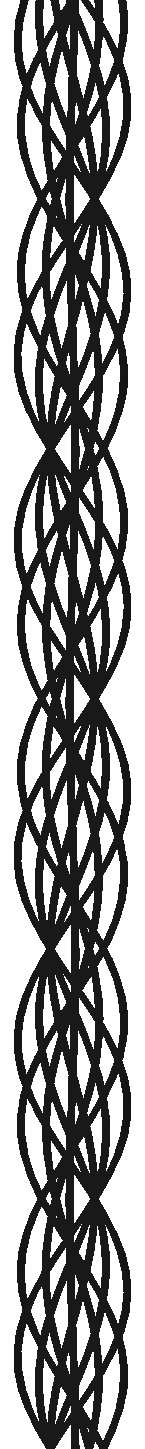} &
    \includegraphics[width=0.25in]{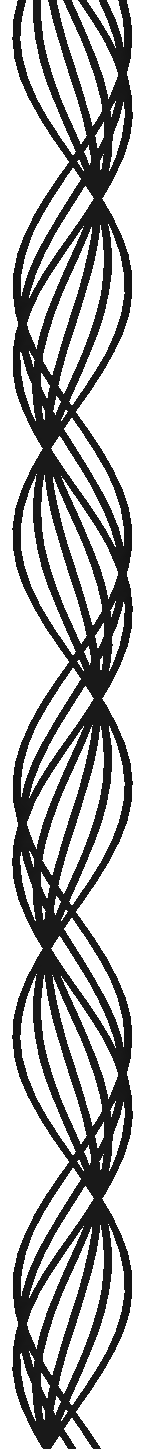} &
    \includegraphics[width=0.25in]{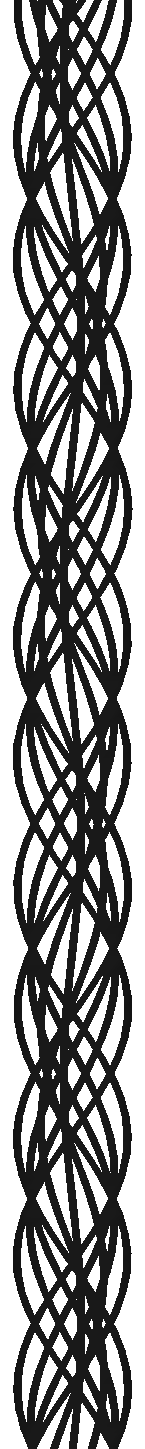}
    \\
    \includegraphics[width=0.25in]{cane/top/N} &
    \includegraphics[width=0.25in]{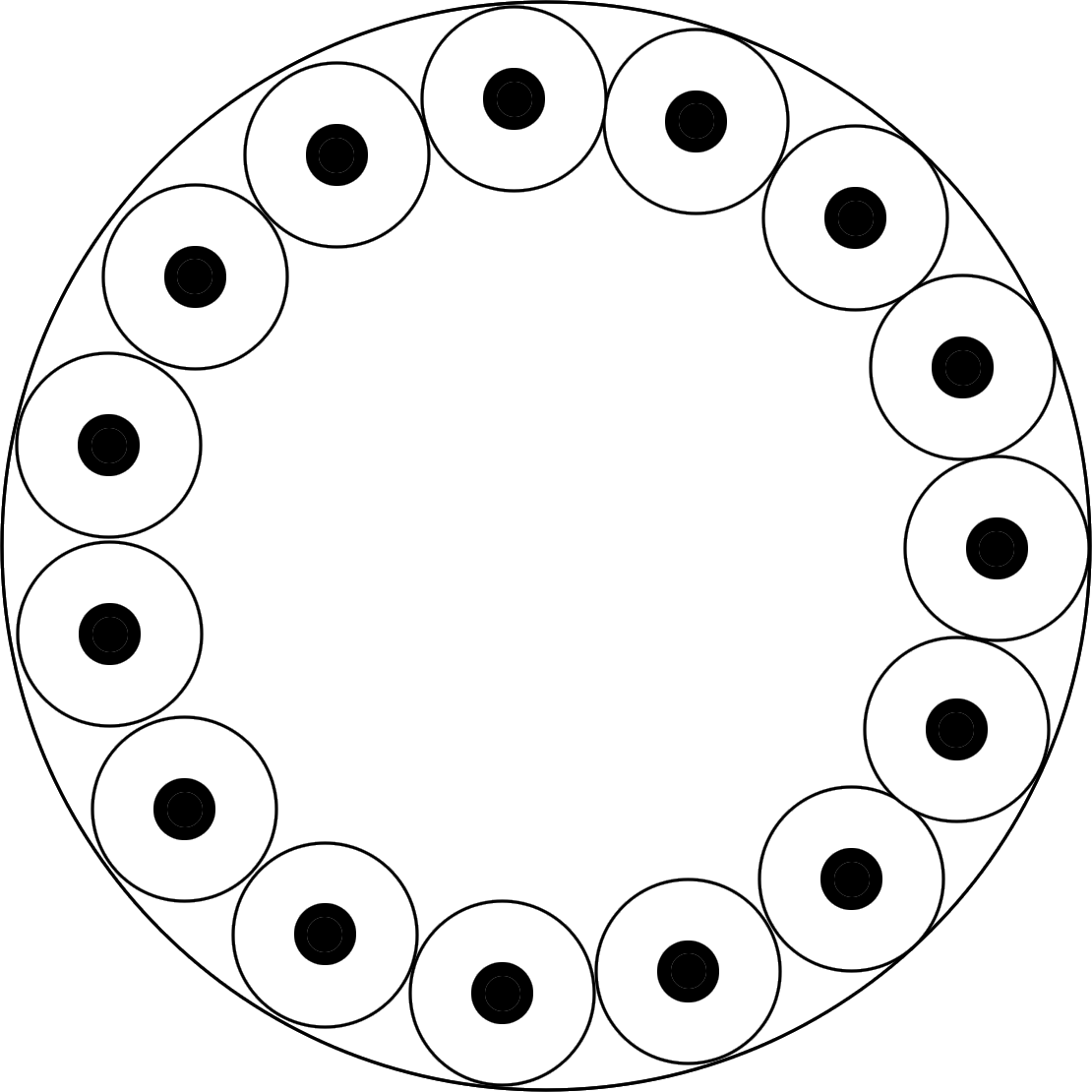} &
    \includegraphics[width=0.25in]{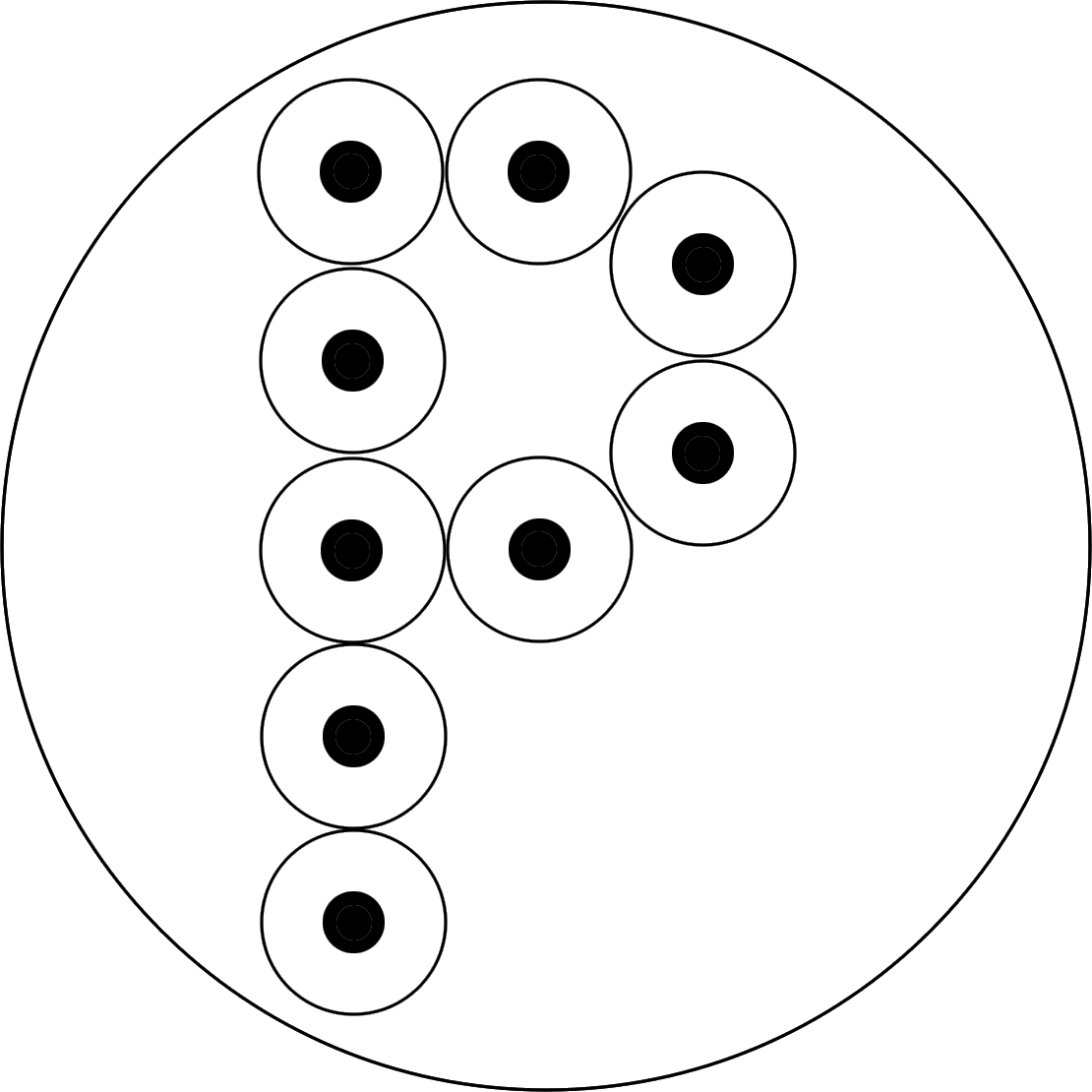} &
    \includegraphics[width=0.25in]{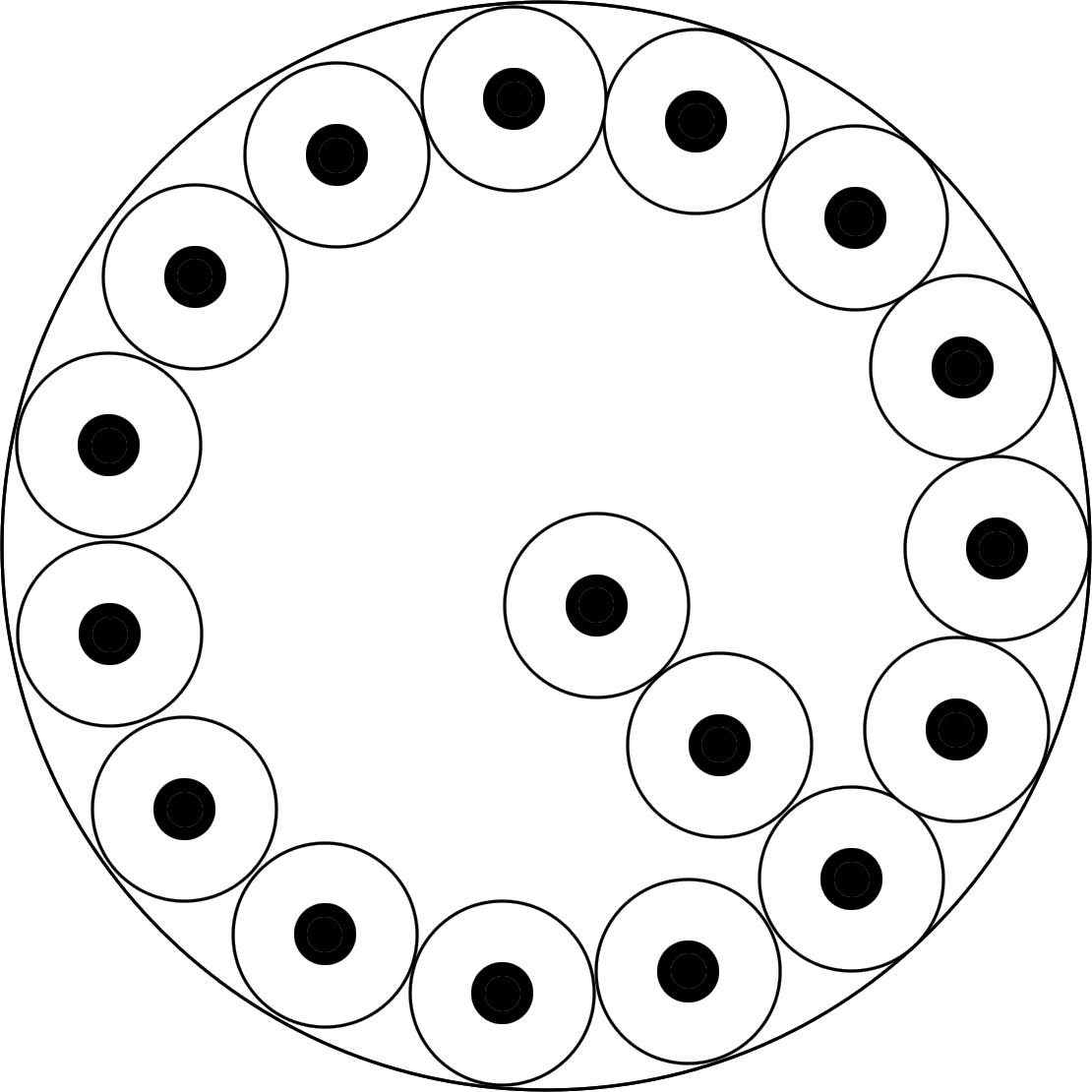} &
    \includegraphics[width=0.25in]{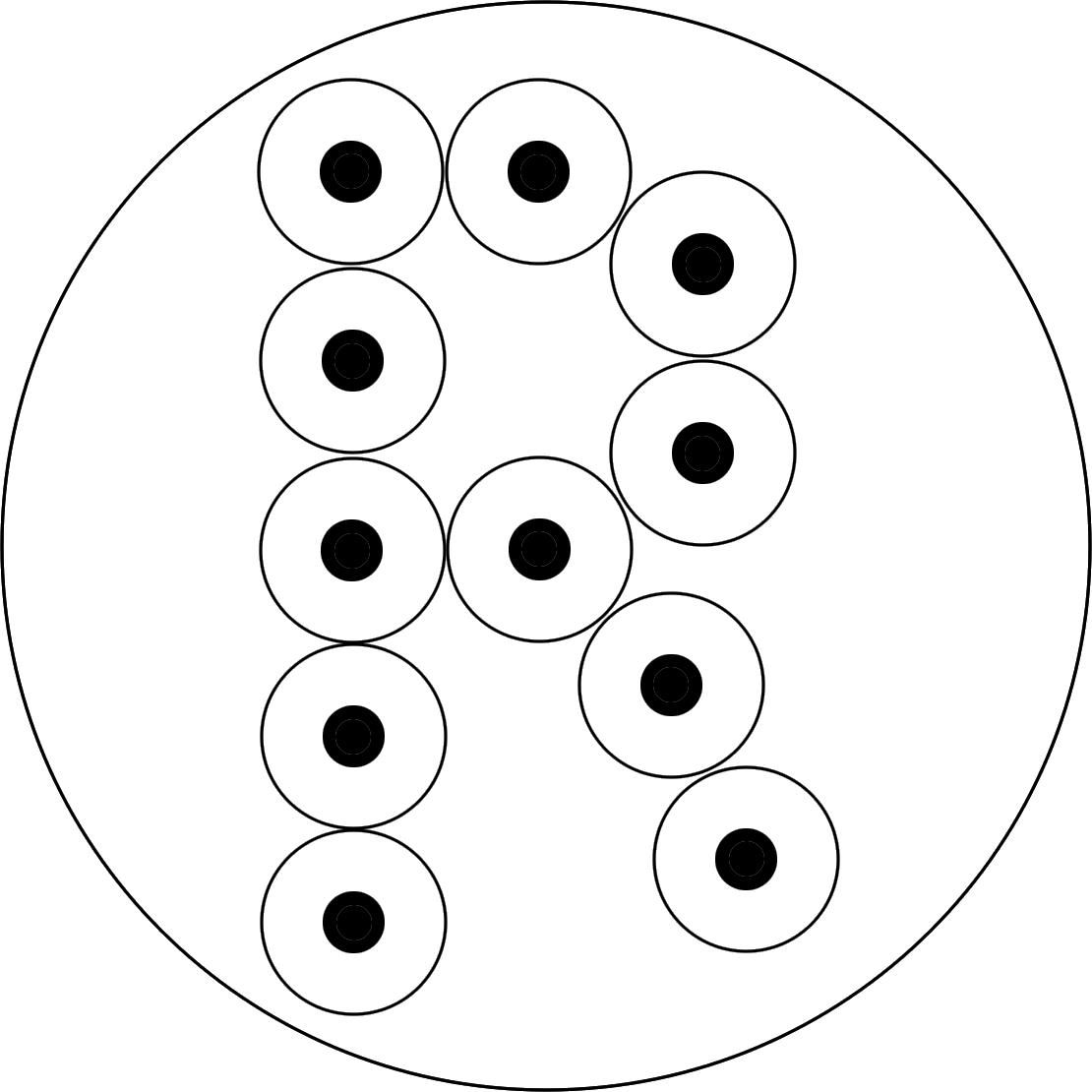} &
    \includegraphics[width=0.25in]{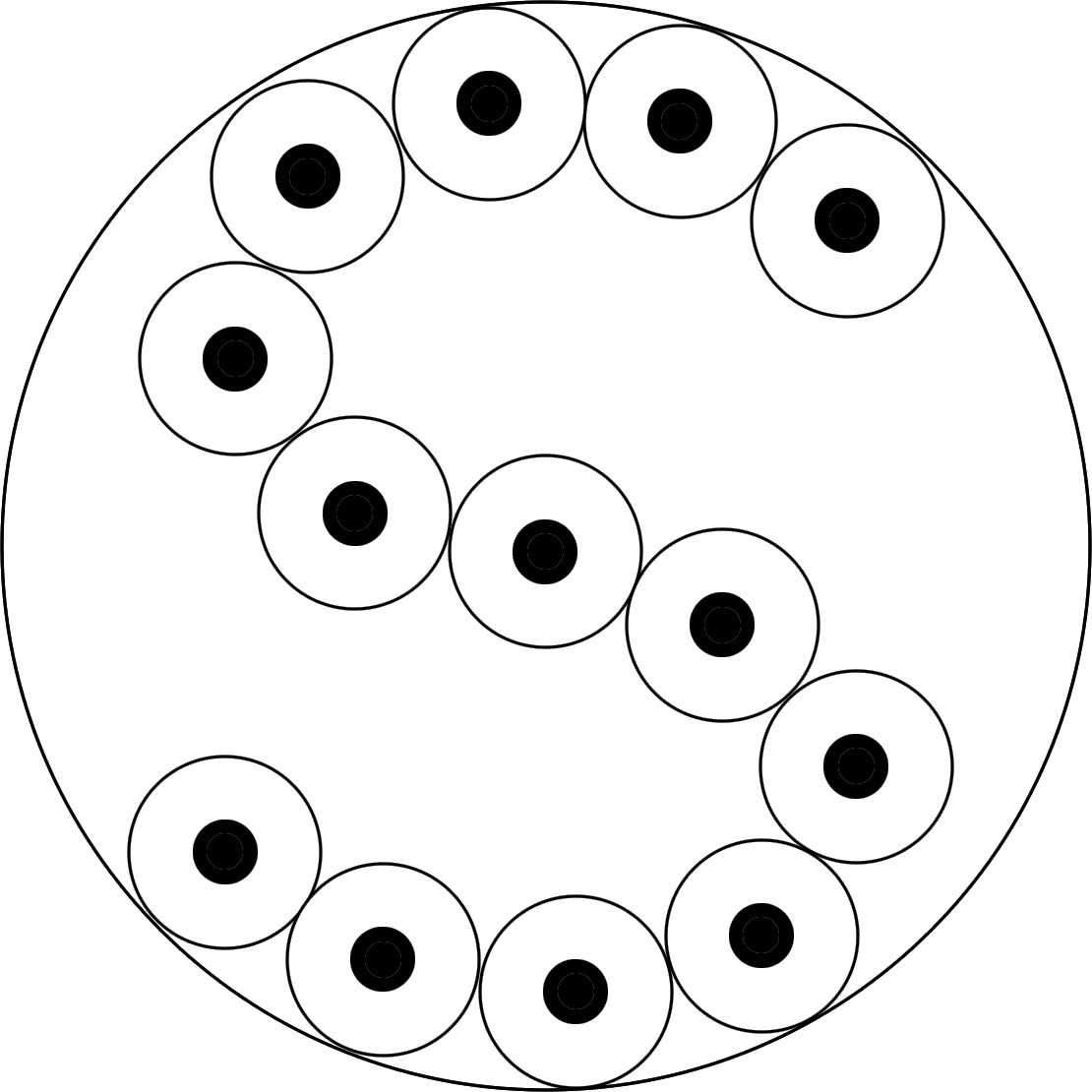} &
    \includegraphics[width=0.25in]{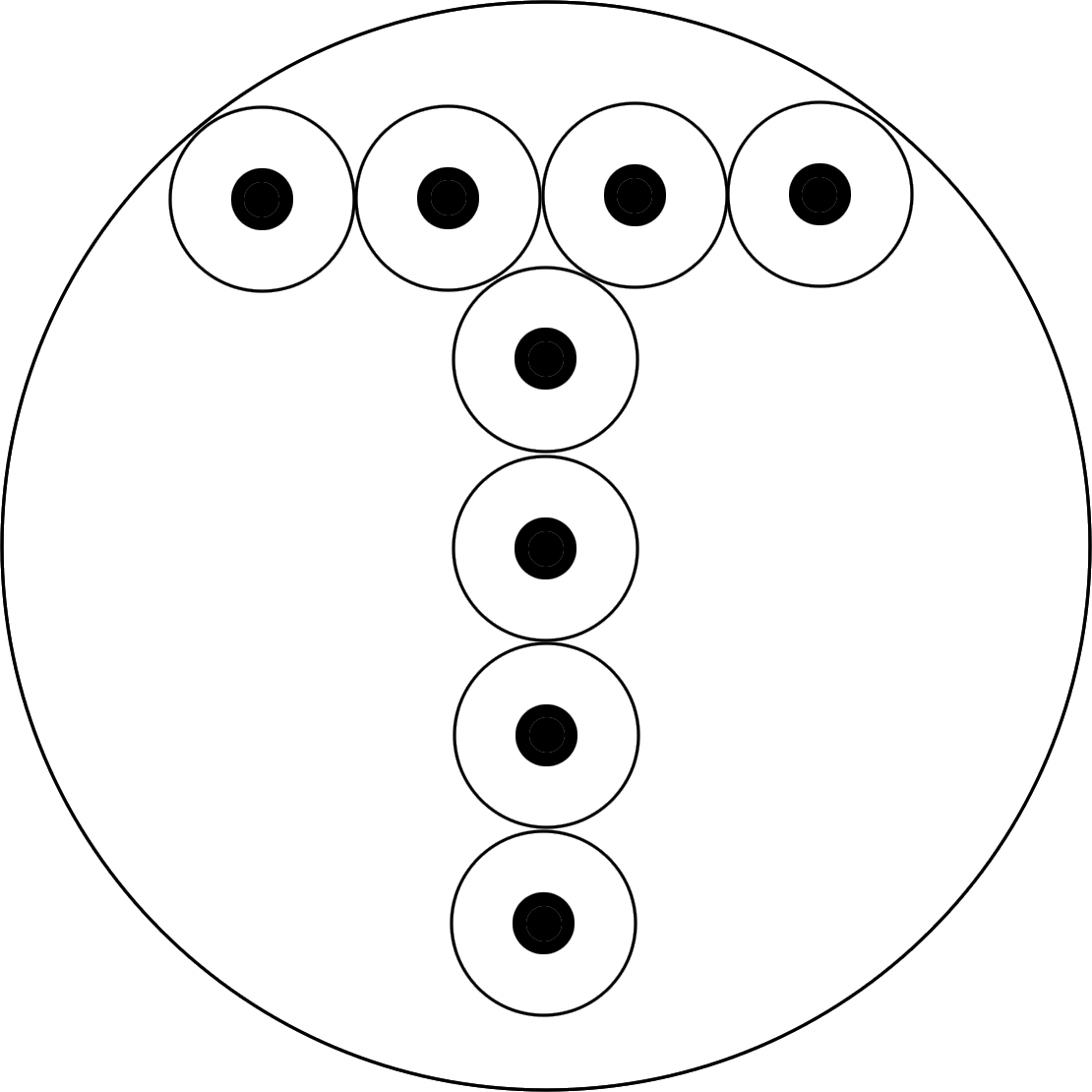} &
    \includegraphics[width=0.25in]{cane/top/U} &
    \includegraphics[width=0.25in]{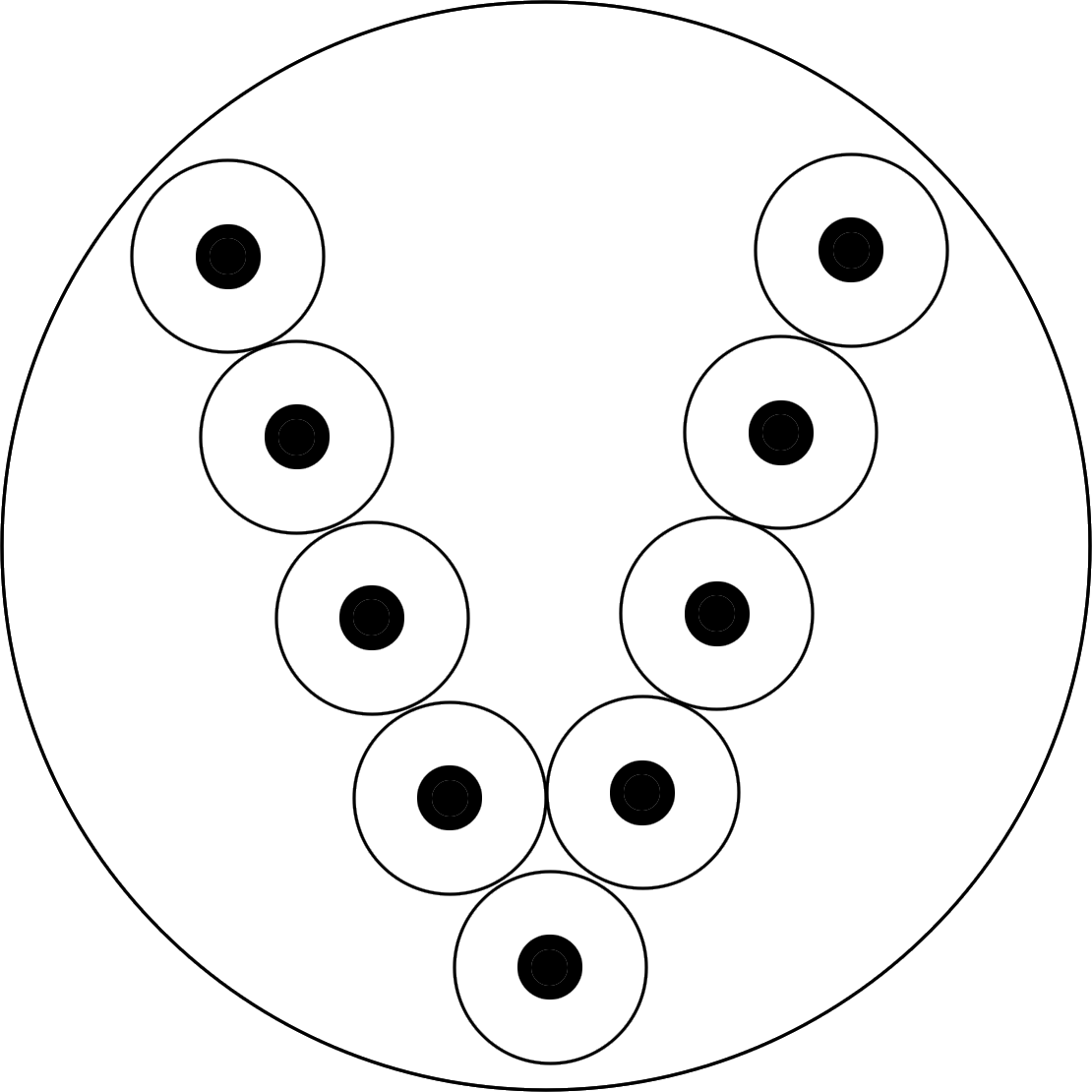} &
    \includegraphics[width=0.25in]{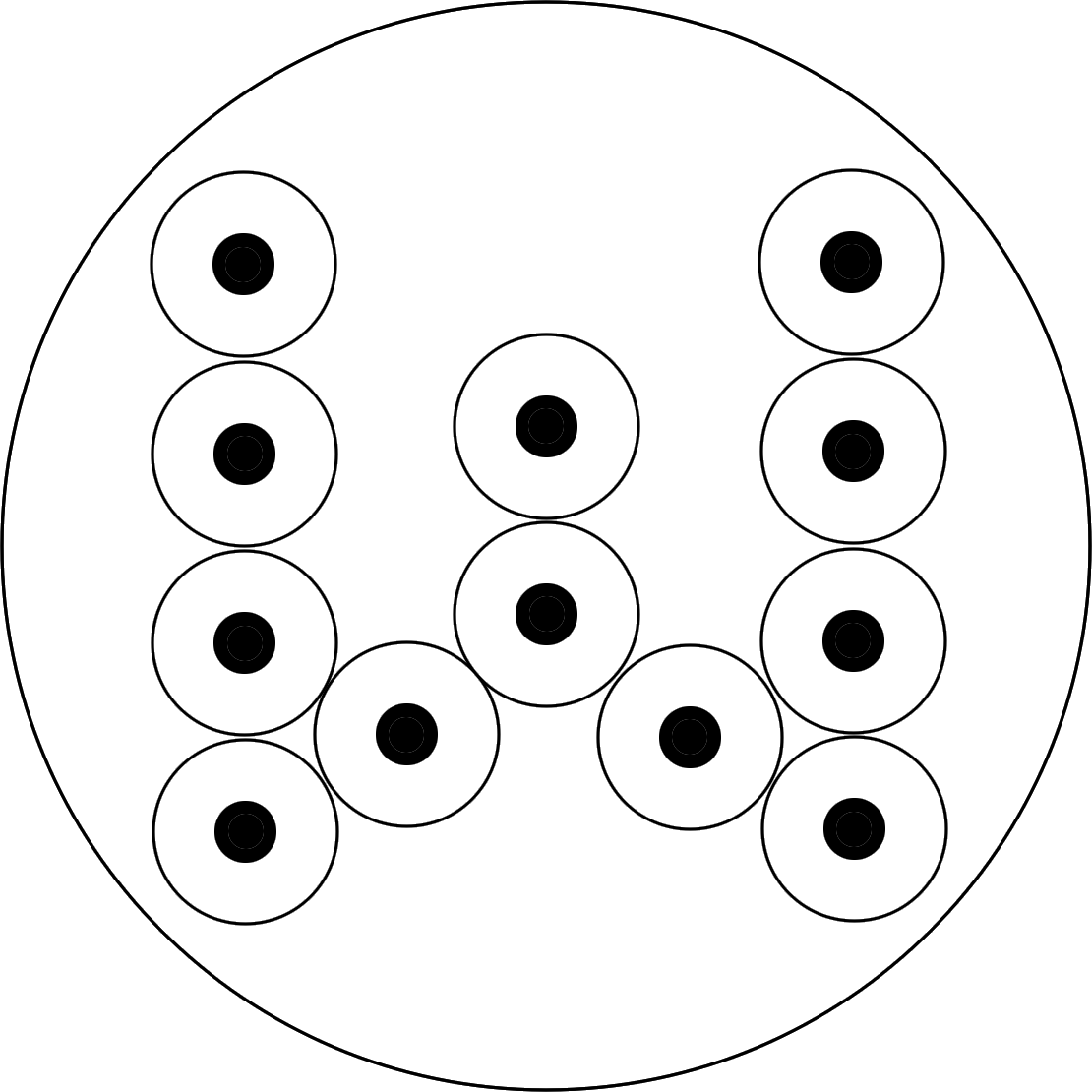} &
    \includegraphics[width=0.25in]{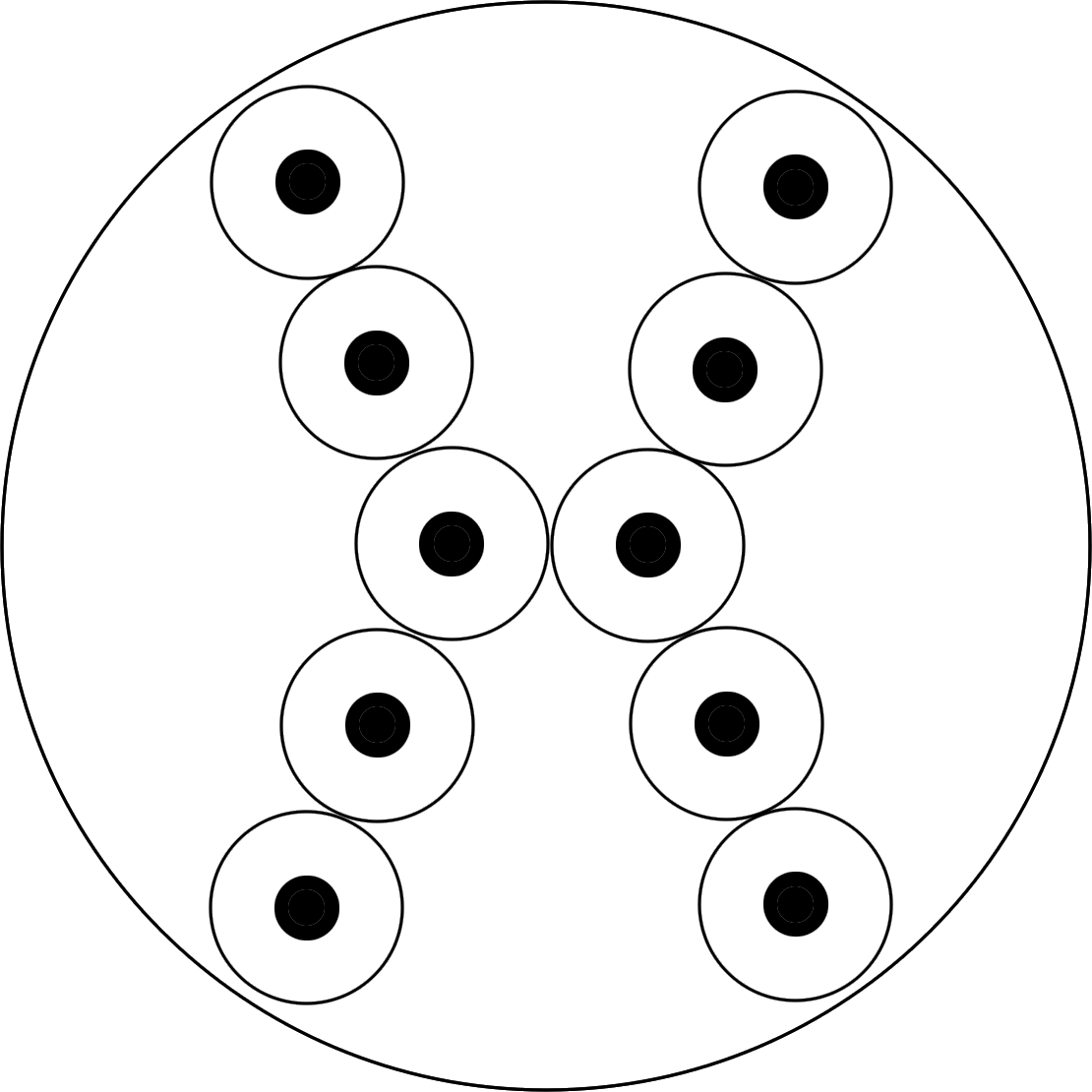} &
    \includegraphics[width=0.25in]{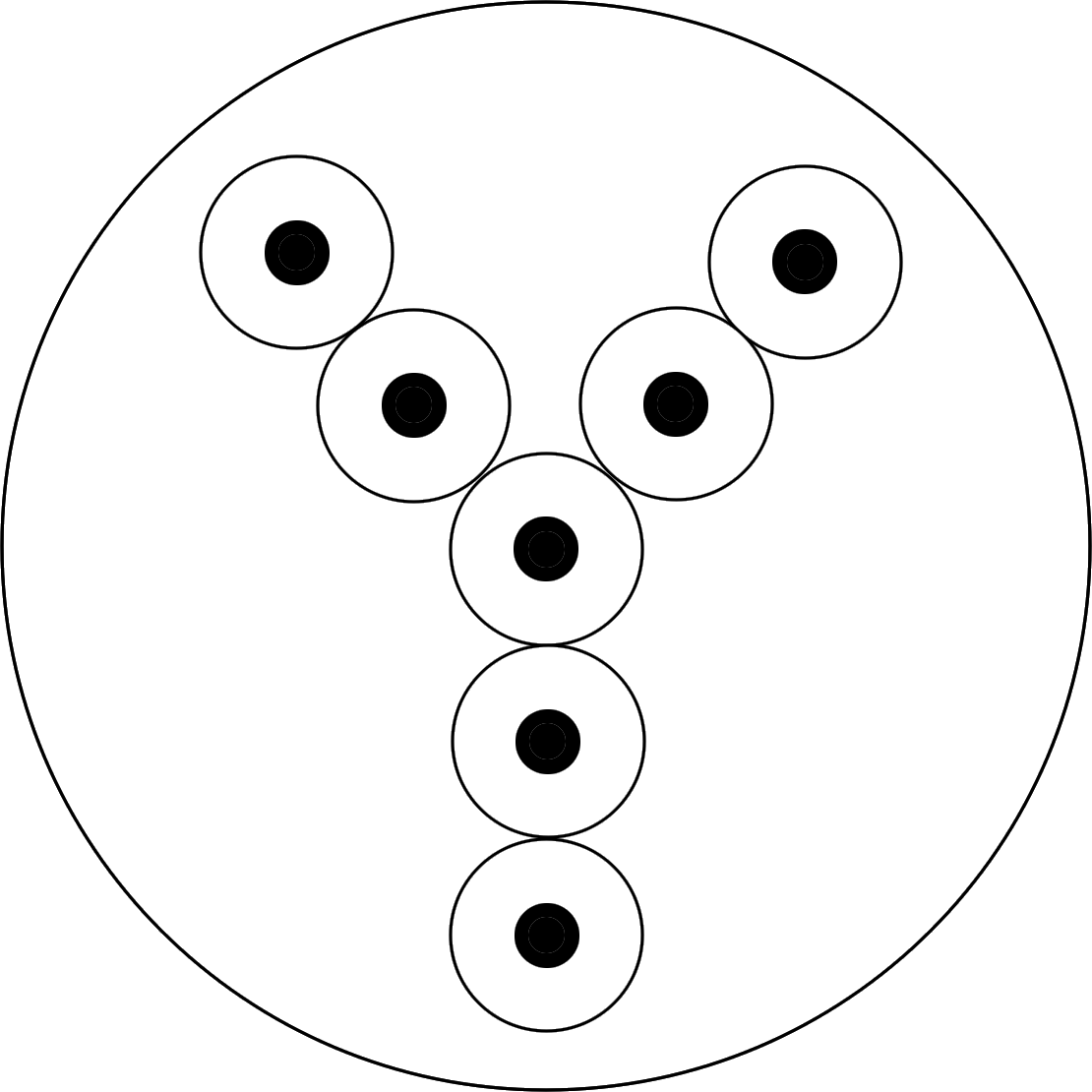} &
    \includegraphics[width=0.25in]{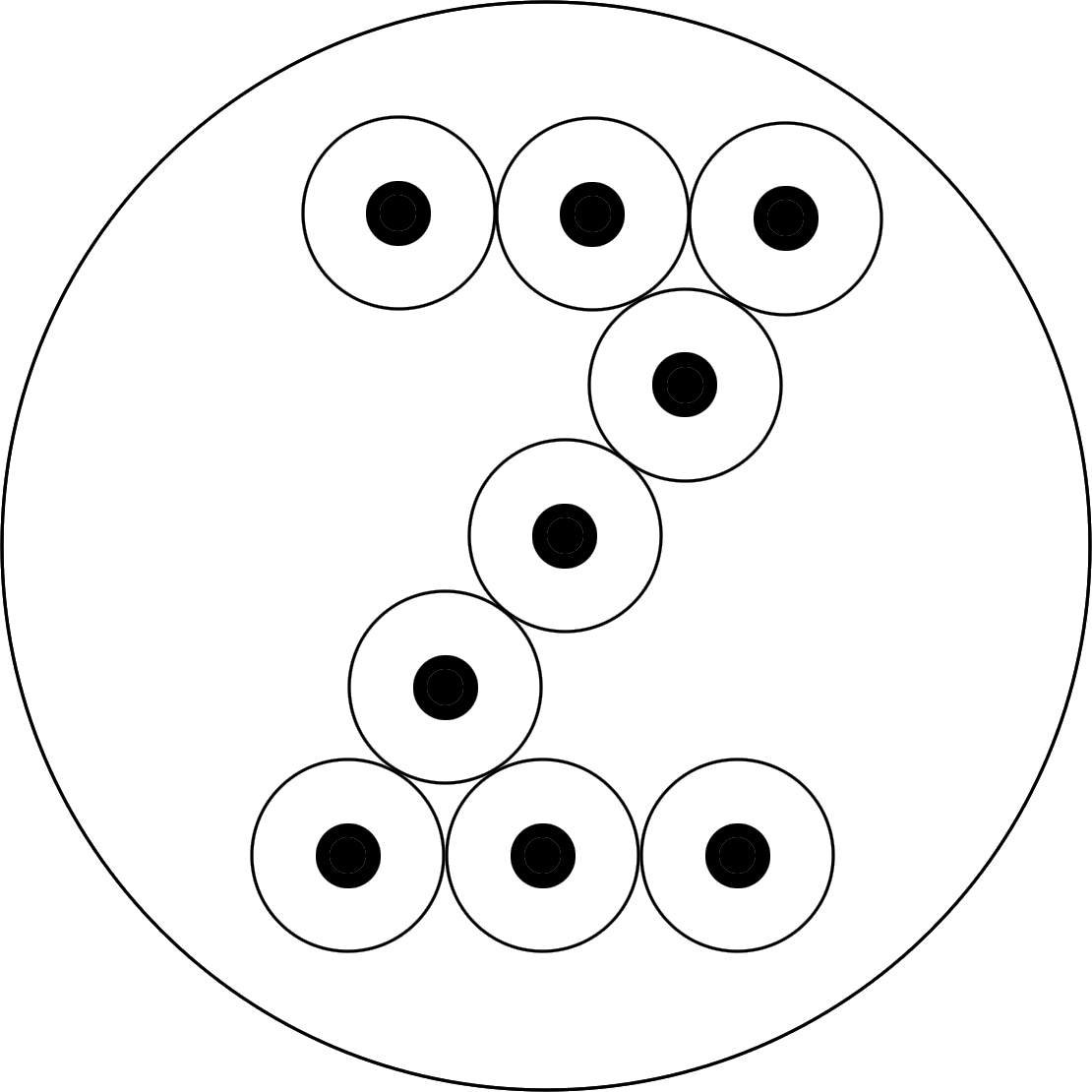}
    \\
    \includegraphics[width=0.25in]{cane/side/N} &
    \includegraphics[width=0.25in]{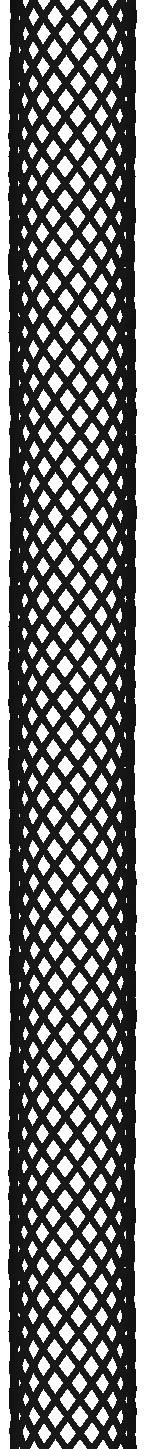} &
    \includegraphics[width=0.25in]{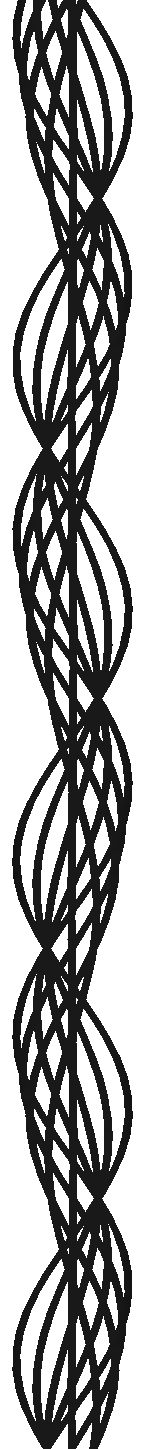} &
    \includegraphics[width=0.25in]{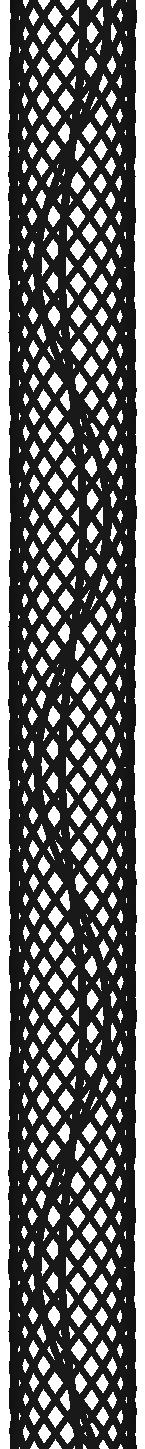} &
    \includegraphics[width=0.25in]{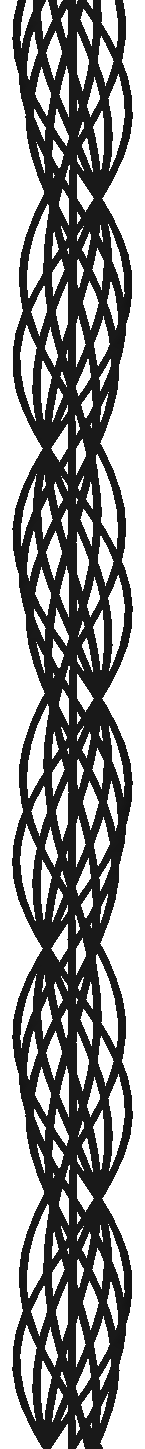} &
    \includegraphics[width=0.25in]{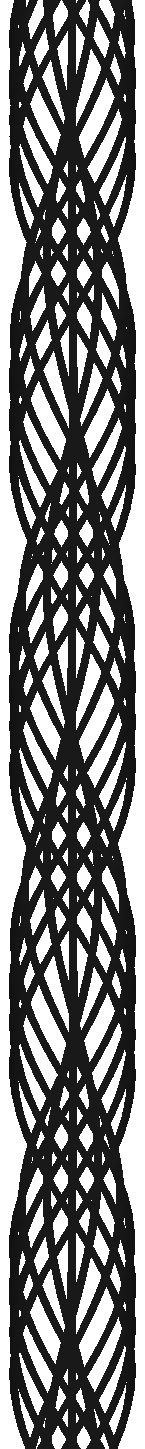} &
    \includegraphics[width=0.25in]{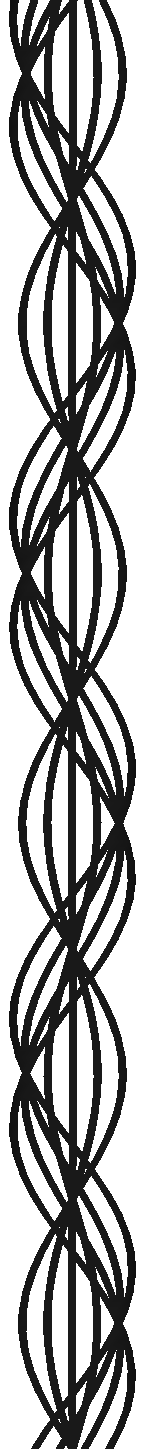} &
    \includegraphics[width=0.25in]{cane/side/U} &
    \includegraphics[width=0.25in]{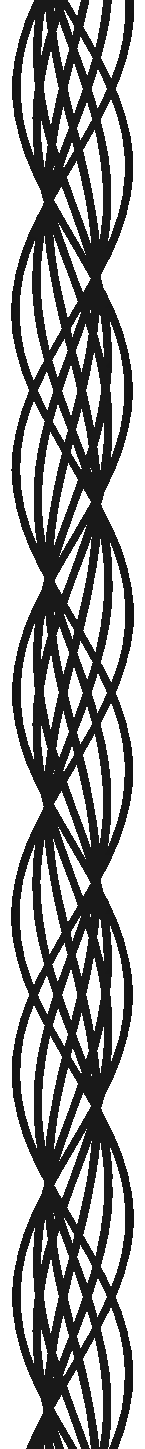} &
    \includegraphics[width=0.25in]{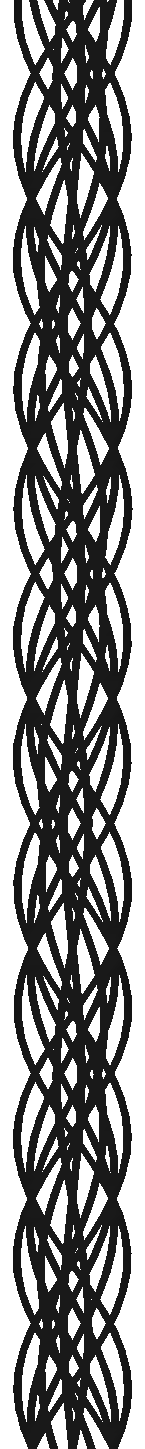} &
    \includegraphics[width=0.25in]{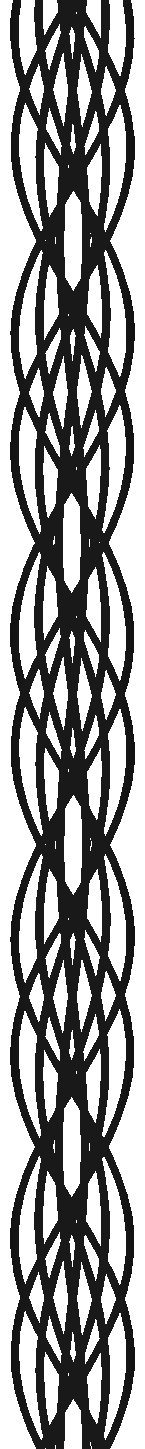} &
    \includegraphics[width=0.25in]{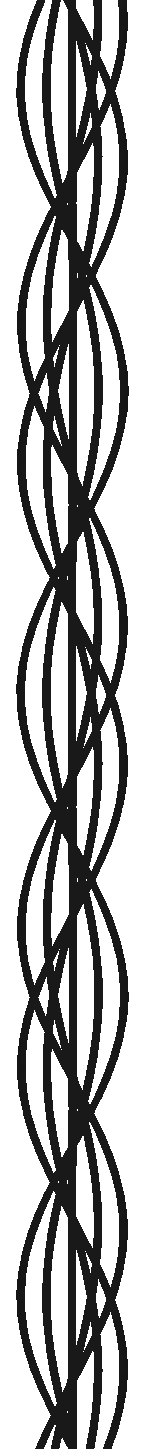} &
    \includegraphics[width=0.25in]{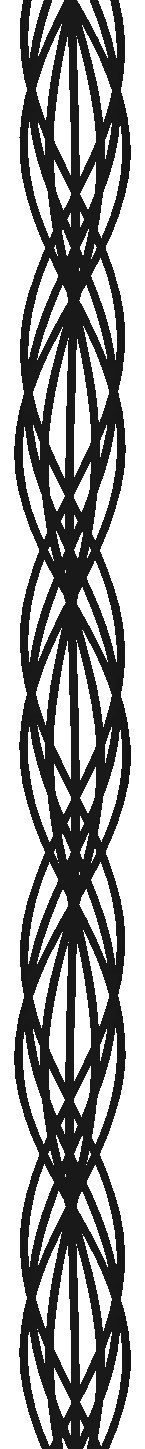}
  \end{tabular}
  \caption{\label{cane font}
    Glass-cane typeface.  Images produced by Virtual Glass.}
\end{figure}

Our glass cane typeface, shown in Figure~\ref{cane font}, consists of two
main fonts.%
\footnote{\url{http://erikdemaine.org/fonts/cane/}}
The easy-to-read font is the top (cross-section) view of each
cane.  The puzzle font is the resulting twisted cane from the side view.
We have designed the canes to be distinguishable from each other, though
some letters require some careful analysis to reverse-engineer.
The letters I and O are classic cane designs.

Another artist and friend, Helen Lee, made her own glass cane font using
Virtual Glass \cite{Helen-font}.

\section{Glass Squishing}

One aspect of glass blowing not currently captured by our Virtual Glass
software is the ability to ``squish'' components of glass together.
This action is a common technique for combining multiple glass structures,
in particular when designing elaborate glass cane.  To model this phenomenon,
we need a physics engine to simulate the idealized behavior of glass under
``squishing''.  But how exactly does glass behave when squished together?

\begin{figure}
  \centering
  \subfloat[Line art, before squishing]{\includegraphics[width=\linewidth]{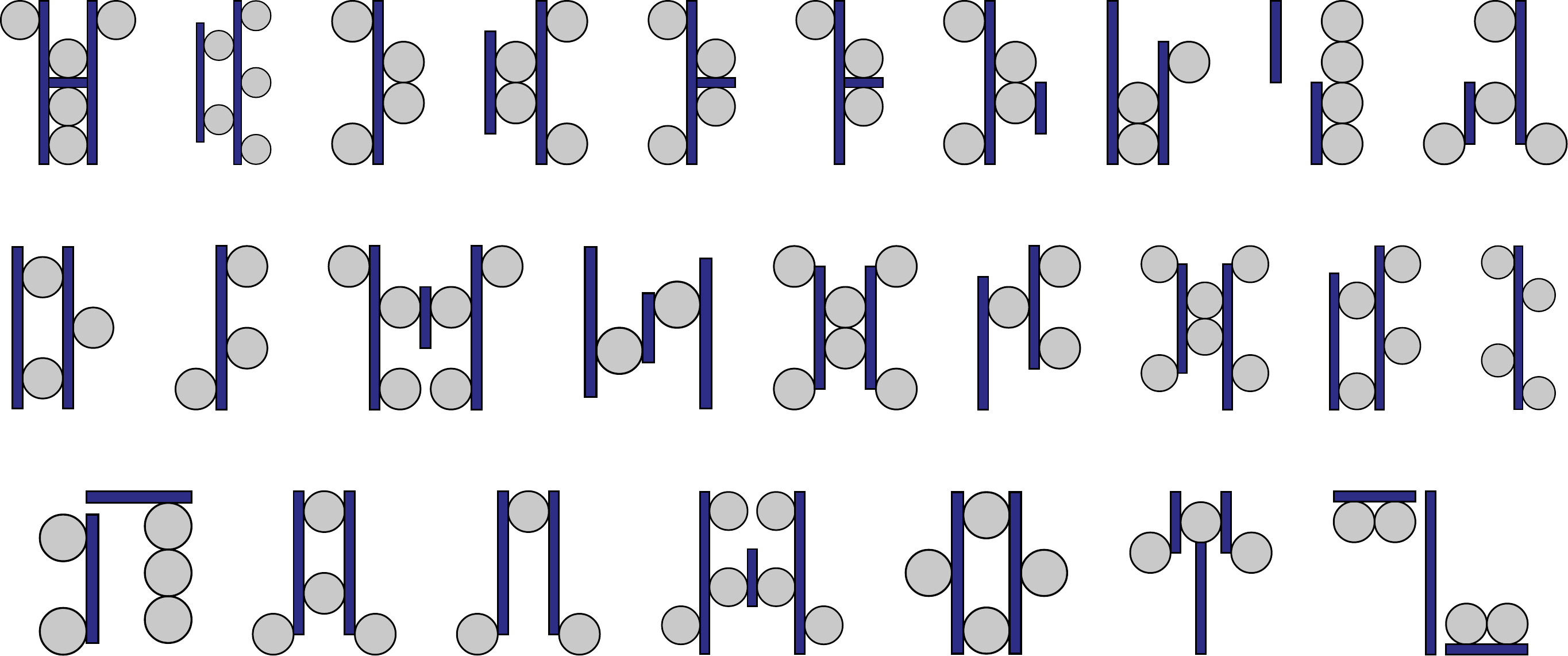}}

  \subfloat[Line art, after squishing]{\includegraphics[width=0.8\linewidth]{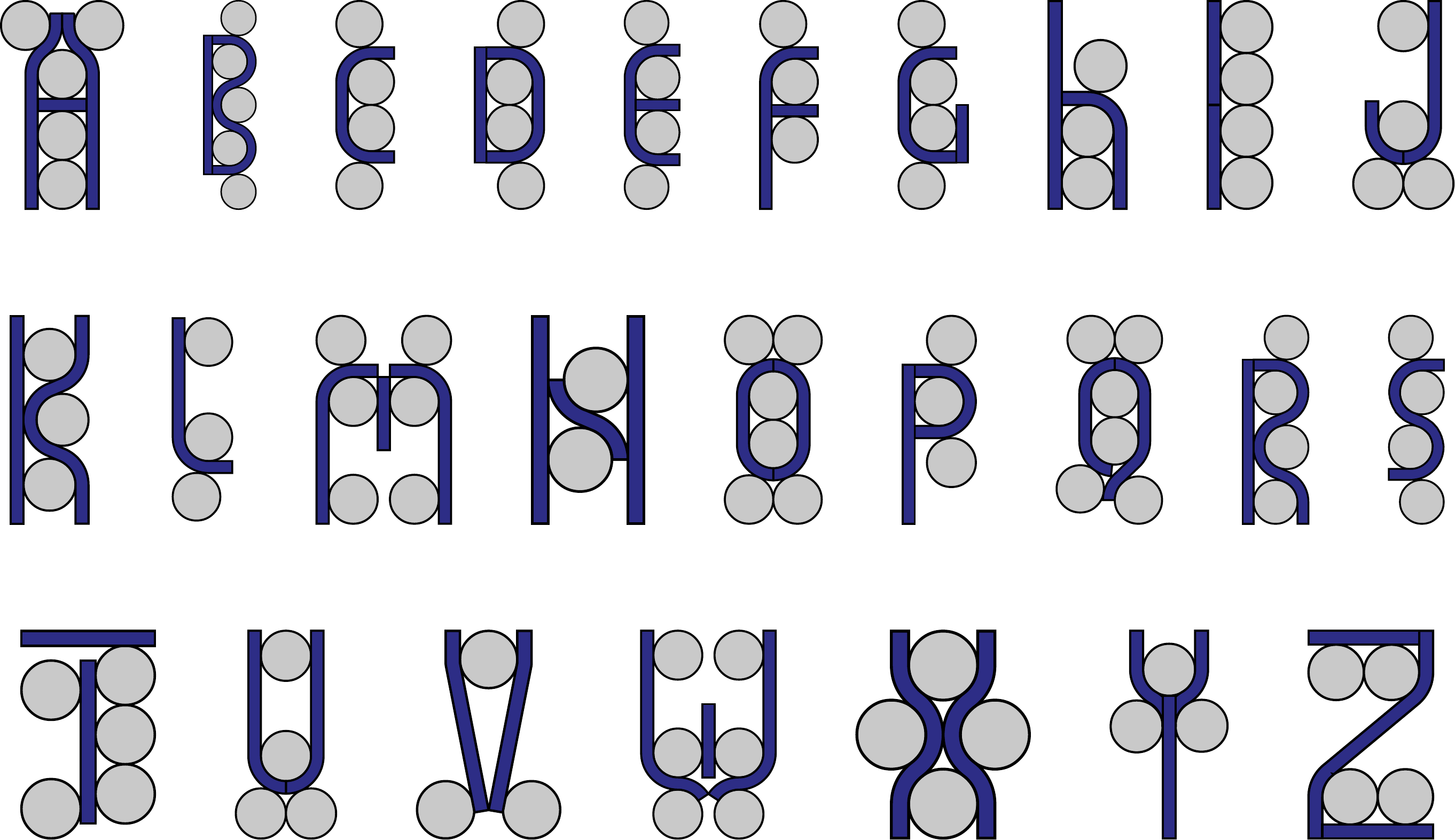}}%
  \caption{Glass-squishing typeface.}
  \label{squish font}
\end{figure}

\begin{figure}
  \centering
  \includegraphics[width=0.48\linewidth]{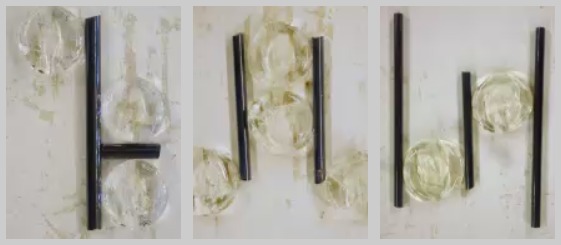}\hfill
  \includegraphics[width=0.48\linewidth]{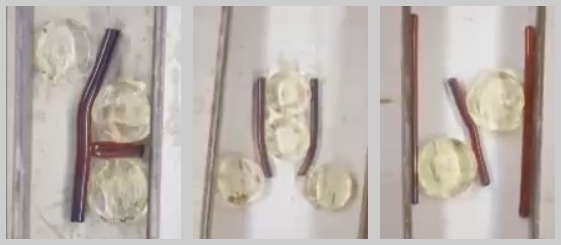}

  \includegraphics[width=0.48\linewidth]{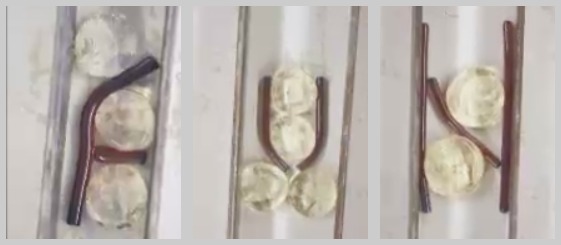}\hfill
  \includegraphics[width=0.48\linewidth]{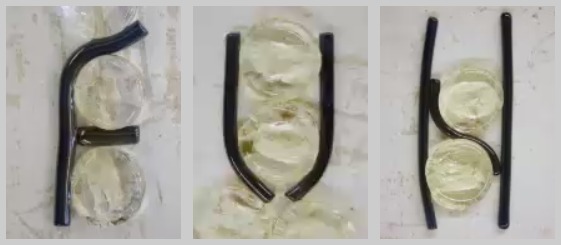}

  \caption{Frames from the video font rendering of F-U-N.}
  \label{squish video}
\end{figure}

To better understand this physical behavior, we designed a glass-squishing
typeface during a 2014 residency at Penland School of Crafts.
As shown in Figure~\ref{squish font}, we designed arrangements of simple
glass components---clear disks and opaque thin lines/cylinders---that,
when heated to around $1400^\circ$F and squished between two vertical steel
bars, produce any desired letter.  The typeface consists of five
main fonts: photographs of the arrangements before and after squishing,
line drawings of these arrangements before and after squishing, and
video of the squishing process.
The ``before'' fonts are puzzle fonts,
while the ``after'' fonts are clearly visible.
The squishing-process font is a rare example of a video font,
where each glyph is a looping video.
Figure~\ref{squish video} shows stills from the video for the letters F-U-N.
See the web app for the full experience.%
\footnote{\url{http://erikdemaine.org/fonts/squish/}}

Designing the before-squishing glass arrangements required extensive trial and
error before the squished result looked like the intended glyph.  This
experimentation has helped us define a physical model for the primary forces
and constraints for glass squishing in 2D, which can model the cross-section
of 3D hot glass.  We plan to implement this physical model to both create
another video font of line art simulating the squishing process, and
to enable a new category of computer-aided design of blown glass
in our Virtual Glass software.
In this way, we use typeface design to experiment with and inform our
computer science research.

\section{Fixed-Angle Linkages}

Molecules are made up of atoms connected together by bonds, with
bonds held at relatively fixed lengths, and incident bonds held at
relatively fixed angles.  In mathematics, we can model these structures as
\emph{fixed-angle linkages}, consisting of rigid bars (segments)
connected at their endpoints, with specified fixed lengths for the bars
and specified fixed angles between incident bars.
A special case of particular interest is a \emph{fixed-angle chain}
where the bars are connected together in a path, which models the
backbone of a protein.  There is extensive algorithmic research on
fixed-angle chains and linkages, motivated by mathematical models of
protein folding; see, e.g., \cite[chapters 8--9]{Demaine-O'Rourke-2007}.
In particular, the literature has studied \emph{flat states} of fixed-angle
chains, where all bars lie in a 2D plane.

\begin{figure}
  \centering
  \includegraphics[width=\linewidth]{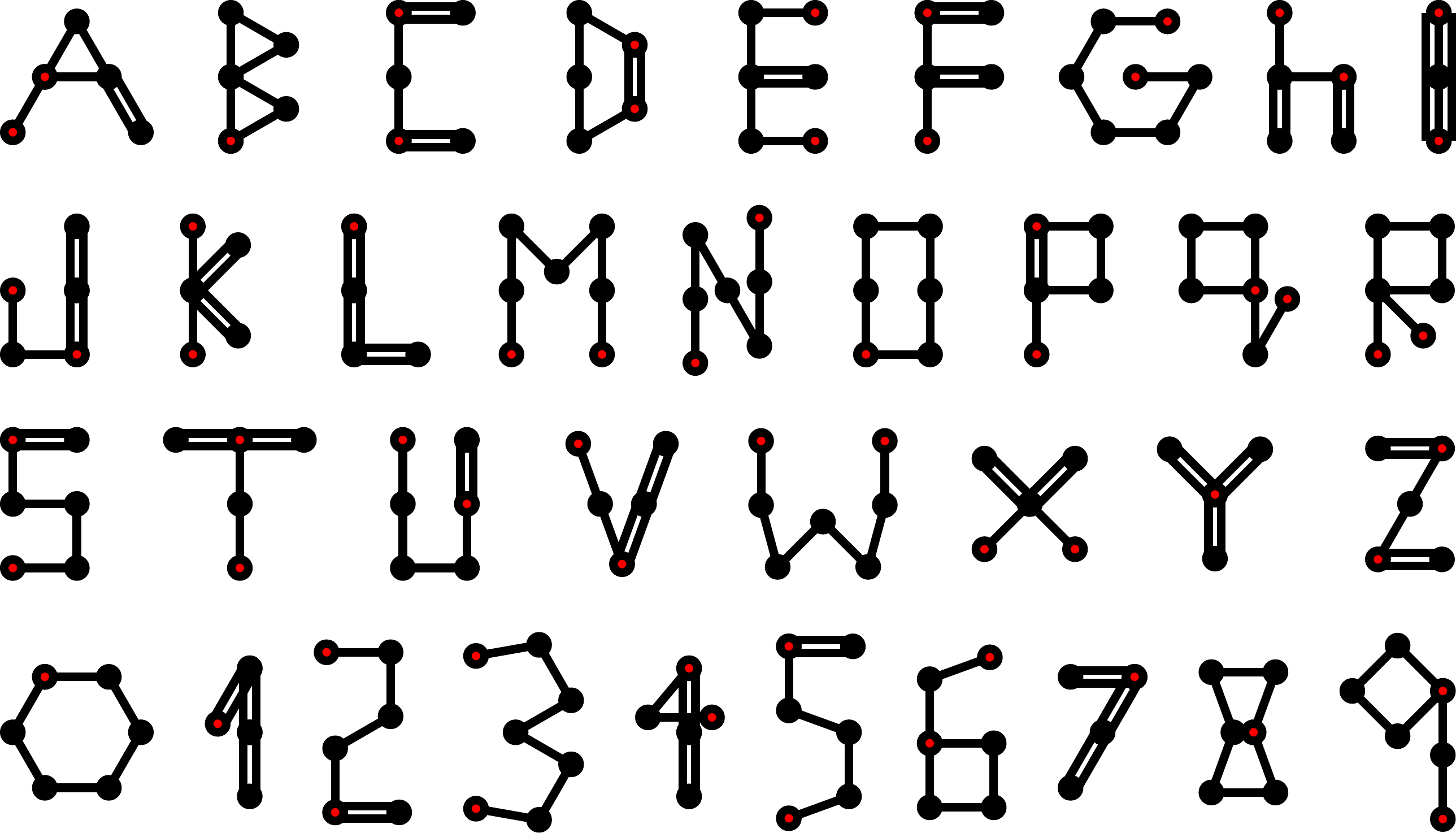}
  \caption{Linkage typeface, from \cite{Demaine-Demaine-2014-linkage-font}.
    Each letter has several glyphs; shown here is the ``correct'' glyph.
    Doubled and tripled edges are spread apart for easier visibility.}
  \label{linkage font}
\end{figure}

Our linkage typeface, shown in Figure~\ref{linkage font}, consists of
a fixed-angle chain for each letter and numeral.  Every fixed-angle
chain consists of exactly six bars, each of unit length.  Hence,
each chain is defined just by a sequence of five measured (convex) angles.
Each chain, however, has many flat states, depending on whether the
convex side of each angle is on the left or the right side of the chain.
Thus, each chain has $2^5 = 32$ glyphs depending on the choice
for each of the five angles.  (In the special cases of zero and $360^\circ$
angles, the choice has no effect so the number of distinct glyphs is smaller.)

%
%\begin{wrapfigure}{r}{1.5in}
%  \centering
%  %\vspace{-2ex}
%  \includegraphics[width=\linewidth]{linkage/fun_variants}
%  \caption{A few random linkage glyphs for F-U-N.}
%  \vspace{-4ex}
%  \label{fun variants}
%\end{wrapfigure}
%

\begin{figure}
  \centering
  %\vspace{-2ex}
  \includegraphics[width=0.3333\linewidth]{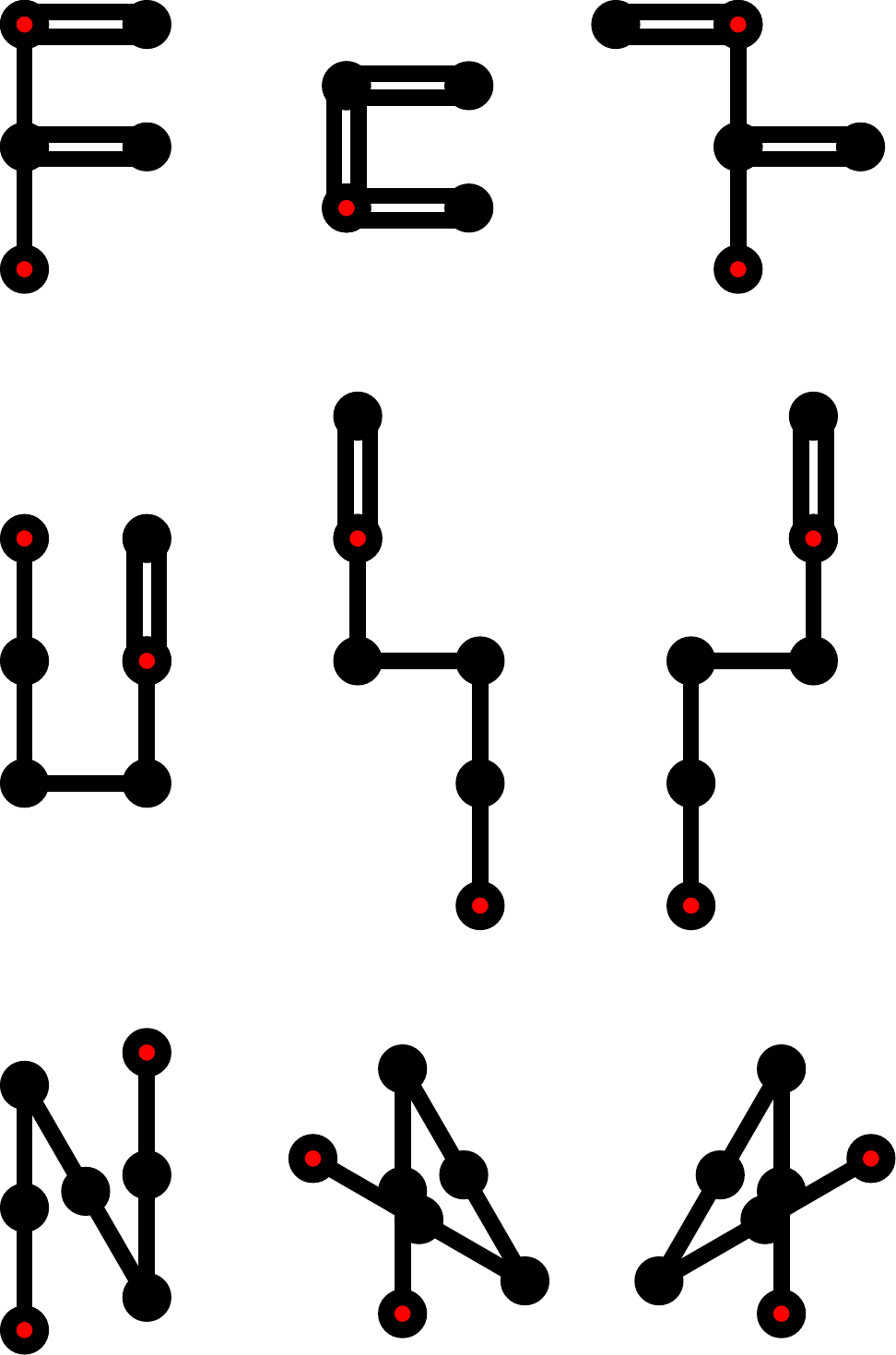}
  \caption{A few random linkage glyphs for F-U-N.}
  %\vspace{-4ex}
  \label{fun variants}
\end{figure}

Thus each letter and numeral has several possible glyphs, only a few of which
are easily recognizable; the rest are puzzle glyphs.
Figure~\ref{fun variants} shows some example glyphs for F-U-N.
We have designed the fixed-angle chains to be uniquely
decodable into a letter
or numeral; the incorrect foldings do not look like another letter or numeral.
The result is a random puzzle font.%
\footnote{\url{http://erikdemaine.org/fonts/linkage/}}
Again we have used this font to design several puzzles
\cite{Demaine-Demaine-2014-linkage-font}.

In addition, there is a rather cryptic puzzle font given just by the
sequence of angles for each letter.  For example, F-U-N can be written as
\hbox{90-0-90-90-0} \hbox{0-180-90-90-180} \hbox{180-30-180-30-180}.

\bibliography{conveyer,dissect,glass,linkage,origami}
\bibliographystyle{alpha}

\end{document}